\documentclass[a4paper,fleqn,usenatbib]{mn2e}

\usepackage{graphicx}
\usepackage{amsmath,amssymb}

\newcommand{\msun}{\rm{M}_\odot}
\newcommand{\nifs}{^{56}\rm{Ni}}
\newcommand{\snx}{SN\,2010X\,}
\newcommand{\snbj}{SN\,2002bj\,}
\newcommand{\snu}{SN\,2015U\,}
\newcommand{\msh}{M_{\rm shell}}
\newcommand{\mcsm}{M_{\rm CSM}}

\title[Bright nickel-free supernovae]{Models of bright nickel-free supernovae from stripped massive stars with circumstellar shells}
\author[Kleiser et al.]{Io Kleiser$^{1}$\thanks{E-mail:ikleiser@caltech.edu}, Daniel Kasen$^{2,3,4}$, Paul Duffell$^{4}$
\\
$^{1}$Department of Astronomy, California Institute of Technology, Pasadena, CA 91125.\\
$^{2}$Lawrence Berkeley National Laboratory, 1 Cyclotron
  Road, Berkeley, CA 94720.\\
$^{3}$Department of Physics, University of California,
  Berkeley, CA 94720.\\
  $^{4}$Department of Astronomy, University of California,
  Berkeley, CA 94720.\\}

\begin{document}

\maketitle

\begin{abstract}
The nature of an emerging class of rapidly fading supernovae (RFSNe)---characterized by their short-lived light curve duration, but varying widely in peak brightness---remains puzzling. Whether the RFSNe arise from low-mass thermonuclear eruptions on white dwarfs or from the core collapse of massive stars is still a matter of dispute. We explore the possibility that the explosion of hydrogen-free massive stars could produce bright but rapidly fading transients if the effective pre-supernova radii are large and if little or no radioactive nickel is ejected. The source of radiation is then purely due to shock cooling. We study this model of RFSNe using spherically symmetric hydrodynamics and radiation transport calculations of the explosion of stripped stars embedded in helium-dominated winds or shells of various masses and extent. We present a parameter study showing how the properties of the circumstellar envelopes  affect the dynamics of the explosion and can lead to a diversity of light curves. We also explore the dynamics of the fallback of the innermost stellar layers, which might be able to remove radioactive nickel from the ejecta, making the rapid decline in the late time light curve possible. We provide scaling relations that describe how the duration and luminosity of these events depend on the supernova kinetic energy and the mass and radius of the circumstellar material.

\end{abstract}

\section{Introduction } \label{s:intro}

The population of observed supernovae (SNe)  is growing swiftly as high-cadence surveys fill regions of observational phase space that were previously much less  accessible. 
Among the peculiar objects found are a class of rapidly fading supernovae (RFSNe) with peak luminosities ranging widely from sub-luminous to brighter than ``normal" SNe. Well known single objects include \snbj \citep{poznanski10}, \snx \citep{kasliwal10}, and \snu \citep{shivvers16}, but studies of the larger population have also emerged \citep[e.g.,][]{drout14, arcavi16}. The progenitor systems and explosion mechanisms of RFSNe these events remain in dispute.

RFSNe exist in what is currently the shortest-timescale region of optical observational parameter space, with rise and decline times  lasting days to weeks. 
If these transients are interpreted as  powered by centrally concentrated radioactive $\nifs$, the total ejected mass must be small 
($\sim 0.1~\msun$, assuming a constant opacity) so as to produce a short effective diffusion time.  
Several theoretical models may produce such ejecta, for example the thermonuclear detonation of a helium shell atop a white dwarf  \citep[a ``point Ia supernova",][]{bildsten07, shen10}, the explosion of a highly stripped massive star \citep{tauris15}, or a core collapse supernova experiencing heavy fallback \citep{moriya10}.

However, low-mass $\nifs$ powered models likely cannot  explain many of the RFSNe. The light curves of many observed events show no noticeable late-time ``tail" indicating a continuing input of decay  energy (although incomplete trapping of the radioactive $\gamma$-rays could perhaps explain this behavior). Moreover, some objects, such as \snbj and \snu, are so bright that simple analytic estimates lead to the unphysical inference that the mass of $\nifs$ must be larger than the total ejecta mass. 
For such reasons, \citet{drout14} conclude that many of the RFSNe are likely powered by shock energy rather than radioactivity.
 
Previous modeling by \cite{kleiser14} has shown that some RFSNe like SN2010X could be explained by the explosion of a hydrogen-poor star with a relatively large radius ($\sim 20~R_\odot$). The ejected mass of radioactive isotopes was assumed to be small, such that the luminosity was powered by diffusion of the shock deposited energy. The model light curves declined rapidly due to recombination in the cooling ejecta (composed of helium or carbon/oxygen) which reduced the opacity and led to a rapid depletion of the thermal energy, similar to the end of the plateau in
Type~IIP SN.   Dim transients of this sort had been studied in the SNIb models of \citet{dessart11}. 
 
To produce a bright RFSN from shock cooling requires a progenitor star with a radius much greater than the few ${\rm R}_\odot$ found in
stellar evolution models of hydrogen-stripped stars \cite{crowther07}.
\cite{kleiser14} suggested that the effective presupernova star radius  may be increased due to envelope inflation of mass loss
just prior to explosion.  Strong mass-loss episodes could arise due to binary interaction \citep{chevalier12} or dynamics driven by nuclear burning \citep{quataert12, smith16}. Indeed, the spectra of Type Ibn SN \citep[e.g.][and citations therein]{pastorello15,pastorello16} and of \snu\ provide direct evidence for a hydrogen-poor circumstellar medium (CSM) around some massive star explosions.

In this paper, we pursue the shock cooling model for RFSN by carrying out a parameter study of the dynamics and shock cooling light curves of supernova exploding into an extended, hydrogen poor CSM.
 In \S \ref{s:analytics}, we provide simple analytic  scalings for how the interaction dynamics and resulting light curve should depend on physical parameters such as the mass and radius of the CSM shell. In \S \ref{s:methods}, we describe a pipeline to model the 1D hydrodynamics of the interaction and the subsequent light curves. In \S \ref{s:results}, we show how different shell parameters affect the dynamics and the possibility of fallback. We present light curves for nickel-free and nickel-rich ejecta profiles, and we explore how Ralyeigh-Taylor mixing effects may effect the results. Finally, \S \ref{s:discussion} contains discussion of our results and their implications for our understanding of RFSNe and the possible outcomes of stellar evolution that could produce such peculiar objects.

\section{Analytics}  \label{s:analytics}
We first present simple analytic scalings that can be used to estimate the  properties of interacting SN. As an idealized model, we consider the case of homologously
expanding SN ejecta running into a stationary CSM shell or wind.  Although the interaction with the CSM will generally occur before the  stellar ejecta has had time to establish homology,  our hydrodynamical models (see \S \ref{s:results}) indicate that the post-shock velocity structure of the exploded star is approximately linear in radius. We therefore assume the ejecta velocity at radius $r$ and time $t$ is $v = r/t$ and describe the ejecta  structure with a broken power law profile \citep{chevalier89} in which the density in the outer layers (above a transition velocity $v_t$) is
\begin{equation}
\rho_{\rm ej} \propto \frac{M_{\rm ej}}{v_t^3 t^3} \biggl(\frac{r}{v_t t}\biggr)^{-n}\,\,,
\end{equation}
where $v_t \propto  (E_{\rm exp}/M_{\rm ej})^{1/2}$, and $M_{\rm ej}$ is the ejecta mass and $E_{\rm exp}$ the  energy of the explosion.

Interaction with the (nearly) stationary CSM will decelerate the ejecta and convert its kinetic energy into thermal energy. By conservation of momentum, the mass of ejecta that can be significantly decelerated in the interaction is of order the total mass of the CSM. 
For the power-law density profile, the  ejecta mass 
above some velocity coordinate $v_0 > v_t$ is
\begin{equation}
M(v_0) =  \int_{v_0}^{\infty} 4 \pi r^2  \rho_{\rm ej}(r) dr \propto \frac{4 \pi }{n-3}  M_{\rm ej}\,\,,
\left(\frac{ v_0}{v_t} \right)^{3-n}
\end{equation}
which assumes $n > 3$.  Setting $M(v_0) \sim \mcsm$ (where $\mcsm$ is the total CSM mass) implies 
that the velocity coordinate above which the ejecta is slowed by the interaction is
\begin{equation*}
v_0 \propto v_t   \left( \frac{M_{\rm ej}}{\mcsm} \right)^\frac{1}{n-3}\,\,.
\end{equation*}
The ejecta kinetic energy contained in the layers above $v_0$ is
\begin{eqnarray}
{\rm KE}(v_0) &=&  \int_{v_0}^{\infty}  \frac{1}{2} \rho_{\rm ej} v^2  4 \pi r^2 dr  
\propto
 M_{\rm ej} v_t^2 \left( \frac{v_0}{v_t} \right)^{3-n}
\end{eqnarray}
which suggests that the energy thermalized in the interaction should scale as
\begin{equation}
E_{\rm th,0} \propto {\rm KE}(v_0) \propto  M_{\rm ej}v_t^2\biggl( \frac{\mcsm}{M_{\rm ej}} \biggr)^\frac{n-5}{n-3}\,\,.
\label{eq:Eth0}
\end{equation}
For $n=8$, for example, the  energy thermalized scales as $(\mcsm/M_{\rm ej})^{3/5}$.

The thermalization of the ejecta kinetic energy will occur over the timescale for the ejecta to accelerate the CSM.
To estimate the interaction timescale we follow the self-similar arguments  of \citep{chevalier92} and assume that the CSM has a power-law density structure of the form
\begin{equation}
\rho_{\rm CSM}(r) \propto \frac{\mcsm}{R_{\rm CSM}^3} \biggl(\frac{r}{R_{\rm CSM}}\biggr)^{-s}\,\,,
\end{equation}
where $R_{\rm CSM}$ is the outer radius of CSM and $s < 3$.
In a self-similar interaction, the ejecta and CSM densities  maintain a constant ratio at the contact discontinuity, 
$\rho_{\rm ej}(r_c)/\rho_{\rm CSM}(r_c) = C$, with $C$ a constant.
This implies that $r_c$, the radius of the contact discontinuity between the ejecta and CSM, evolves as \citep{chevalier92}
\begin{equation}
r_c(t) = t^{\frac{n-3}{n-s}} \biggl[ \frac{M_{\rm ej}}{\mcsm} \frac{R_{\rm CSM}^{3-s}}{ C v_t^{3-n}}\biggr]^{\frac{1}{n-s}}\,\,.
\end{equation}
Setting $r_c(t) \approx R_{\rm CSM}$ gives an estimate of the time $t_{\rm bo}$ when the forward shock from interaction will breakout of the CSM \citep{harris16}:
\begin{equation}
t_{\rm bo} \approx \frac{R_{\rm CSM}}{v_t} \biggl(\frac{C \mcsm}{M_{\rm ej}}\biggr)^{\frac{1}{n-3}}\,\,.
\end{equation}
The total amount of ejecta kinetic energy thermalized will rise until $t \approx t_{\rm bo}$, then decline as the interaction abates and the system adiabatically expands.  
Because the pressure is radiation dominated (adiabatic index $\gamma=4/3$) the thermal energy after expansion to a radius $R(t)$ is 
\begin{equation}
E_{\rm th}(t) = E_{\rm th,0}\frac{R_{\rm CSM}}{R(t)}  \propto E_{\rm th,0} \left( \frac{t_{\rm bo}}{t} \right)\,\,,
\end{equation}
where $R(t)$ is the radius of the expanding, post-interaction ejecta, and the last equation assumes homologous expansion, $R(t) \sim t$, following the breakout. The thermal energy at time $t$ is then
\begin{equation}
E_{\rm th}(t) \propto R_{\rm CSM} M_{\rm ej}^{1/2}E_{\rm exp}^{1/2}\biggl( \frac{\mcsm}{M_{\rm ej}} \biggr)^\frac{n-4}{n-3}   t^{-1}.
\end{equation}
For the case of $n = 8$, for example, which will approximate the post-shock density structure of our hydrodynamical models, we have
\begin{equation}
E_{\rm th}(t) \propto R_{\rm CSM} E_{\rm exp}^{1/2}\mcsm^{4/5} M_{\rm ej}^{-4/5} t^{-1}.
\label{eq:Etht}
\end{equation}
We will show using hydrodynamical models in \S\ref{s:param} that Equation~\ref{eq:Etht} accurately predicts how the thermal energy content depends on the CSM and ejecta properties. The derivation assumes $\mcsm \lesssim M_{\rm ej}$.

The light curves arising from the interaction are the result of the diffusion of thermal radiation from the shocked region. The opacity $\kappa$ is usually dominated by electron scattering and is constant in ionized regions, but will drop sharply to near zero once
 the temperature drops below the recombination temperature $T_I$.  Scaling
relations for the duration and peak luminosity of thermal supernovae, including the effects of recombination, have been determined by \citet{popov93} and verified numerically by \citet{kasen09}
\begin{equation}
t_\mathrm{sn} \propto  E_{\rm th,0}^{-1/6}M_{\rm diff}^{1/2}R_0^{1/6}\kappa^{1/6}T_I^{-2/3},
\label{eq:t_sn_1}
\end{equation}
\begin{equation}
L_\mathrm{sn} \propto E_{\rm th,0}^{5/6}M_{\rm diff}^{-1/2} R_{0}^{2/3} \kappa^{-1/3}T_I^{4/3}\,\,,
\label{eq:L_sn_1}
\end{equation}
where $M_{\rm diff}$ is the effective amount of mass the photons must diffuse through. We take this to be some combination of $M_{\rm ej}$ and $M_{\rm CSM}$, depending on the distribution of thermal energy among the relative masses. Taking  $R_0 = R_{\rm CSM}$ and  using our Equation~\ref{eq:Eth0} for $E_{\rm th,0}$ gives
\begin{equation}
t_\mathrm{sn} \propto  E_{\rm exp}^{-1/6}
\biggl(\frac{M_{\rm CSM}}{M_{\rm ej}}\biggr)^{\frac{-(n-5)}{6(n-3)}}
M_{\rm diff}^{1/2}
R_{\rm CSM}^{1/6}
\kappa^{1/6}T_I^{-2/3}\,\,,
\label{eq:t_sn}
\end{equation}
\begin{equation}
L_\mathrm{sn} \propto E_{\rm exp}^{5/6} 
\biggl(\frac{M_{\rm CSM}}{M_{\rm ej}}\biggr)^{\frac{5(n-5)}{6(n-3)}}
M_{\rm diff}^{-1/2}
R_{\rm CSM}^{2/3} 
\kappa^{-1/3}T_I^{4/3}\,\,.
\label{eq:L_sn}
\end{equation}

For the purposes of easy comparison to numerical data, we would like to devise simple power laws to describe the dependency of $L_{\rm sn}$ and $t_{\rm sn}$ on the parameters. This is complicated by the $M_{\rm diff}$ factor, but there are limits we can consider. First it is necessary to recognize that the masses change the light curve in two opposing ways: increasing $\frac{M_{\rm CSM}}{M_{\rm ej}}$ increases the amount of available thermal energy to power the light curve, which would increase the peak luminosity and decrease the timescale, according to Equations \ref{eq:t_sn_1} and \ref{eq:L_sn_1}. Meanwhile, the diffusion mass $M_{\rm diff}$ also slows the diffusion of photons out of the ejecta more as it increases, lowering the peak luminosity and increasing the timescale.

In the cases presented here, we hold $M_{\rm ej}$ fixed. One limit is to imagine that the circumstellar mass is small compared to the ejecta mass, so the dependence on $M_{\rm diff}$ goes away. Then the equations become
\begin{equation}
t_\mathrm{sn} \propto  E_{\rm exp}^{-1/6}
M_{\rm CSM}^{\frac{-(n-5)}{6(n-3)}}
R_{\rm CSM}^{1/6}
\kappa^{1/6}T_I^{-2/3}\,\,,
\label{eq:t_sn_a}
\end{equation}
\begin{equation}
L_\mathrm{sn} \propto E_{\rm exp}^{5/6} 
M_{\rm CSM}^{\frac{5(n-5)}{6(n-3)}}
R_{\rm CSM}^{2/3} 
\kappa^{-1/3}T_I^{4/3}\,\,.
\label{eq:L_sn_a}
\end{equation}
In the case of $n = 8$, we then have $t_{\rm sn} \propto M_{\rm CSM}^{-1/10}$ and $L_{\rm sn} \propto M_{\rm CSM}^{1/2}$. If $n=6$, $t_{\rm sn} \propto M_{\rm CSM}^{-1/18}$ and $L_{\rm sn} \propto M_{\rm CSM}^{5/18}$.

This limit essentially assumes the increase in circumstellar mass does not contribute significantly to inhibiting the travel of photons out of the ejecta. Alternatively, we can imagine that the CSM makes up the bulk of the mass available, or that the total mass scales roughly as the CSM mass. In this case, $M_{\rm diff} \propto M_{\rm CSM}$, so
\begin{equation}
t_\mathrm{sn} \propto  E_{\rm exp}^{-1/6}
M_{\rm CSM}^{\frac{-(n-5)}{6(n-3)} + \frac{1}{2}}
R_{\rm CSM}^{1/6}
\kappa^{1/6}T_I^{-2/3}\,\,,
\label{eq:t_sn_b}
\end{equation}
\begin{equation}
L_\mathrm{sn} \propto E_{\rm exp}^{5/6} 
M_{\rm CSM}^{\frac{5(n-5)}{6(n-3)} - \frac{1}{2}}
R_{\rm CSM}^{2/3} 
\kappa^{-1/3}T_I^{4/3}\,\,.
\label{eq:L_sn_b}
\end{equation}
For $n=8$, $t_{\rm sn} \propto M_{\rm CSM}^{2/5}$ and $L_{\rm sn} \propto M_{\rm CSM}^{0}$. For $n=6$, $t_{\rm sn} \propto M_{\rm CSM}^{4/9}$ and $L_{\rm sn} \propto M_{\rm CSM}^{-2/9}$. We will find in \S \ref{s:results} that this last case with $n=6$ appears to fit our numerical results for the light curves most closely.

\section{Methods } \label{s:methods}

We adopt a spherically symmetric framework to model the light curves of hydrogen-poor stars exploding into an extended CSM.  We use the MESA stellar evolution code to model massive stars that have lost their hydrogen envelopes due to heavy mass loss.  At the point of core collapse, we add to the
MESA models a  parameterized  external shell or wind of mass $\mcsm$. We map this progenitor
structure into a 1D hydrodynamics code and explode it by depositing a central bomb of thermal energy. Once the ejecta have neared homologous
expansion, the  structure is  fed into the 
SEDONA radiation transport code to calculate time-dependent light curves and spectra.

\subsection{Progenitor Star Models}

\begin{figure}
\begin{center}
\includegraphics[width=3.5in]{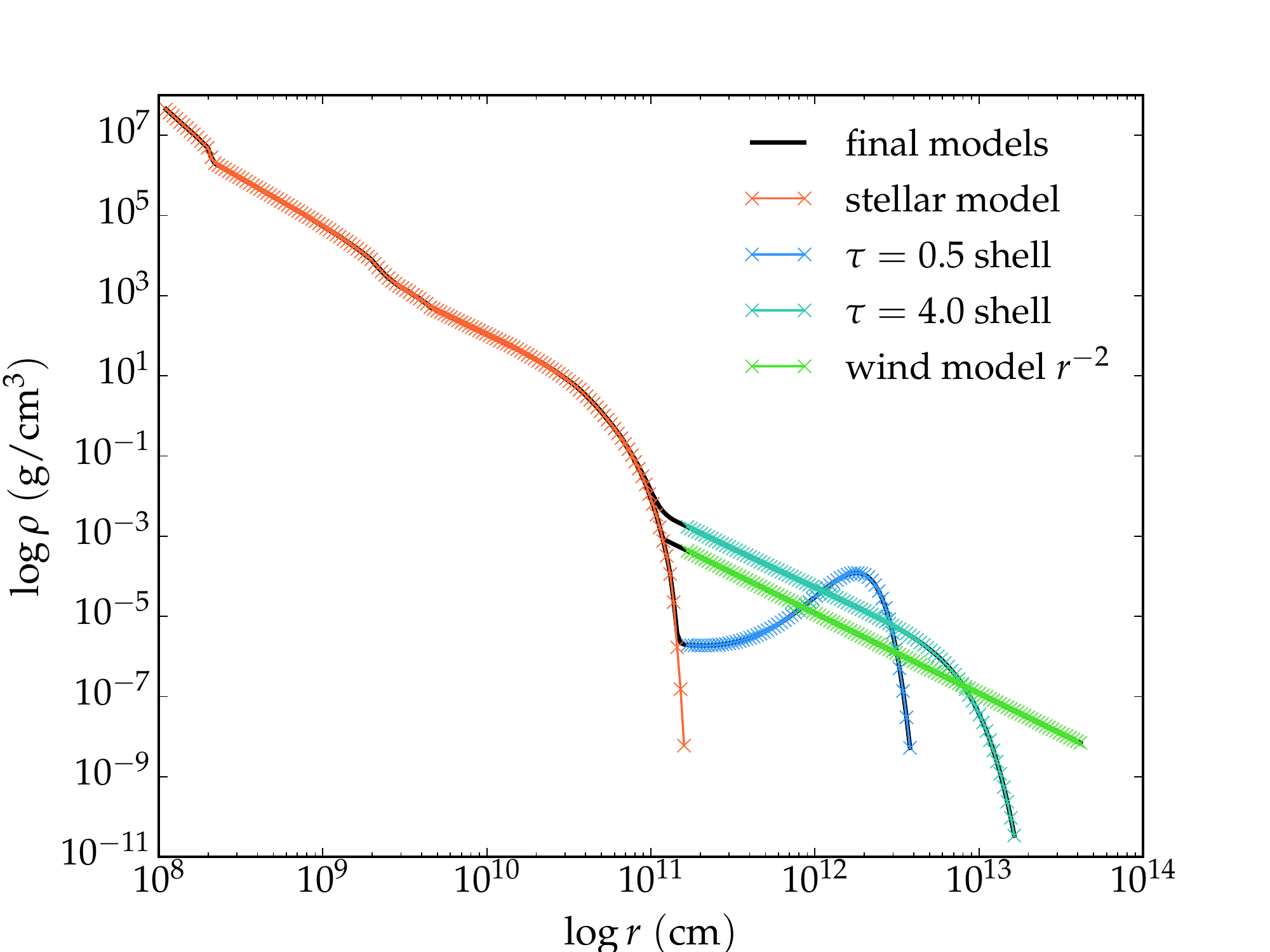}
\end{center}
\caption{Density profile for an example star + shell model. The same stripped MESA star model is used throughout this paper, and different toy shells are constructed around it. The original stellar profile is shown in orange. Blue-green colors show various shell profiles. Two of the shells shown here are Gaussian profiles modified by $r^{-2}$ based on the fact that we assumed a Gaussian $\dot{M}$ whose velocity was constant (see Equation \ref{eq:gauss_shell}) with different values of $\tau$. The third is simply a density profile $\propto r^{-2}$, corresponding to a constant wind prior to explosion. This is essentially the case of infinite $\tau$. Final models are shown in black, with a smooth transition between stellar and shell densities. All shells in this plot have the same amount of total mass. \label{f:shell_profile}}
\end{figure}

\begin{figure}
\begin{center}
\includegraphics[width=3.5in]{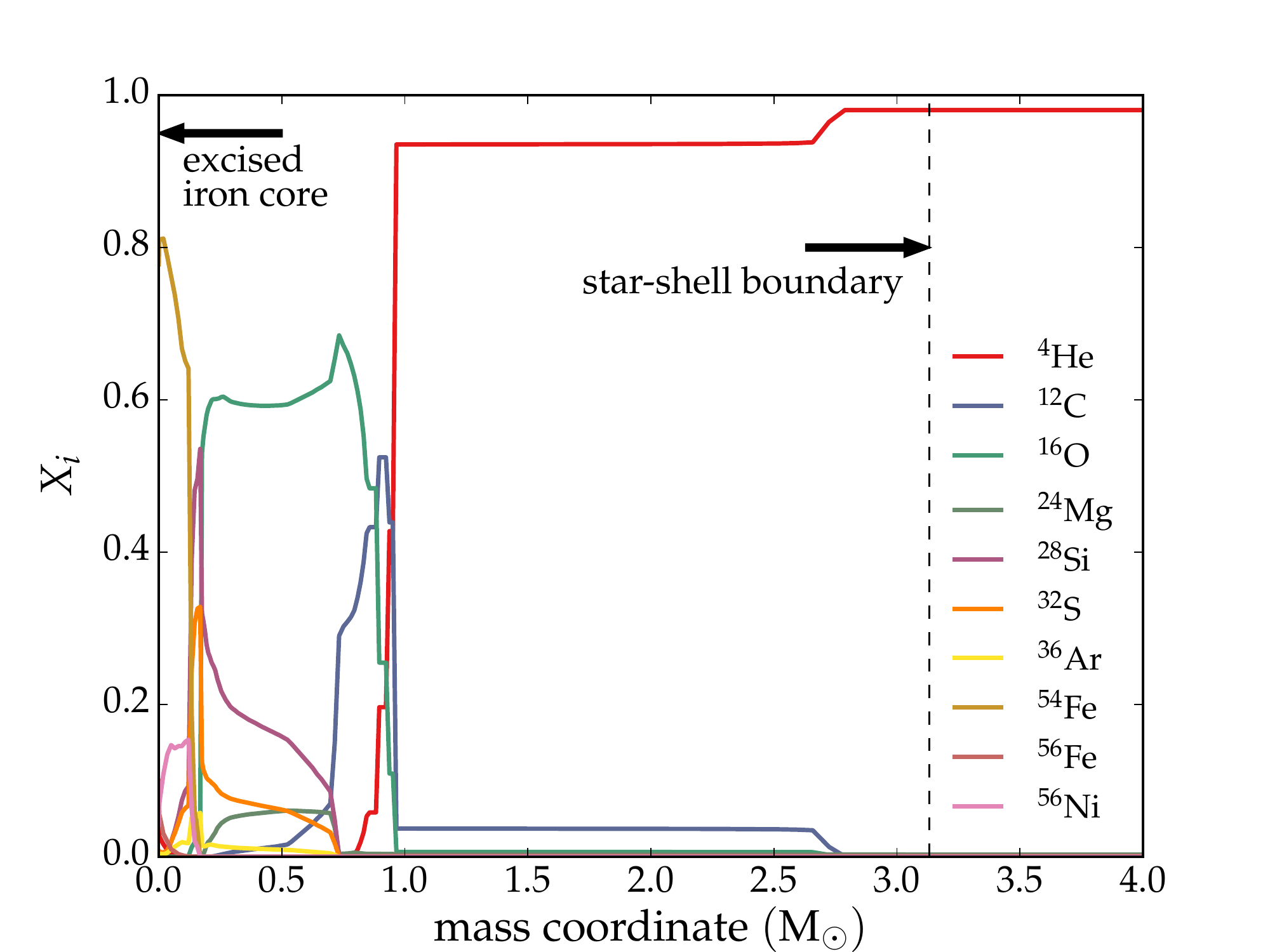}
\end{center}
\caption{Composition plot for an example star + shell model. The iron core has been removed already by cutting out the mass interior to the point where $^{56}{\rm Fe}$ drops below 10\% of the composition. The star used for all runs is the same, and the shell is assumed to have the same abundances as the outermost layer of the star. In this case, all shells are very dominated by $^4 {\rm He}$. The dotted black line indicates where the star ends and the shell begins.\label{f:comp}}
\end{figure}

We use MESA version~7184 to produce a hydrogen-stripped stellar model using a simple artificial mass loss prescription. The prescription is meant to approximate Case B mass transfer to a binary companion, which should be  common among the massive progenitors of Type Ibc SNe \citep[see][]{sana12,smith11}. We use a zero-age main sequence (ZAMS) mass of 20 $\msun$ and evolve the star through hydrogen burning until the surface temperature reaches $T_{\rm eff} = 5000~{\rm K}$, indicating that the radius has expanded significantly.  We then initiate  a constant mass loss at $\dot{M} = 10^{-3}~\msun~{\rm yr}^{-1}$ until a desired final mass is reached, in the present case 5 $\msun$.  This mass loss history qualitatively resembles the more detailed Roche lobe overflow calculations in \citet{yoon10}. Therefore, even though the mass loss prescription is simple, it is similar to the natural loss of a large amount of mass (in this case the entire hydrogen envelope) expected in some systems by Roche lobe overflow. Other or more complex mass loss histories may yield different final stellar structures. 

The MESA model is evolved to the point of iron core collapse. Before exploding the model, we first cut out the remnant based on the point at which $^{56}{\rm Fe}$ drops below 10\% going outward---in our case, the remnant mass is $1.395~\msun$. We then insert an ad-hoc distribution of extended CSM, which
is meant to mock up a heavy mass loss episode in the final days before explosion.  We assume that the CSM mass was lost at a constant velocity,
$v_{\rm CSM} \ll v_{\rm ej}$ with a rate $\dot{M}$ that was Gaussian in time. This leads to a CSM density profile
\begin{equation}
\rho_{\rm CSM} (r) =
 \frac{\mcsm}{ 4 \pi r^2 \Delta r \sqrt{2 \pi} } \exp \biggl[ \frac{-(r - r_{\rm mid})^2}{2 \Delta r}\biggr]\,\,,
\label{eq:gauss_shell}
\end{equation}
where $r_{\rm mid}$ and $\Delta r$  are free parameters specifying, respectively, the peak and the width of the Gaussian. 
For a constant mass rate and wind velocity, $\Delta r = v_{\rm CSM} \tau$ where $\tau$ is the standard deviation of the Gaussian and can be used as a measure of the duration of the
mass loss episode.  For large values of $\tau$, the CSM resembles that of a constant $\dot{M}$ wind with a 
$1/r^2$ density profile. We chose here $v_{\rm CSM} = 100~{\rm km~s^{-1}}$. While the value of $v_{\rm CSM}$ would be interesting in the context of understanding the nature and mechanism of the mass loss, here the actual quantity is of little consequence for the light curves and spectra since the velocity of the ejecta is so much greater.

Figure \ref{f:shell_profile}  shows the density profile of the progenitor star model with a few
different distributions of CSM.  Figure \ref{f:comp} shows the composition of a progenitor model. We assume that the CSM 
composition is homogenous and equal to that at the surface of the stellar model, which is helium-dominated. 

Our parameterized progenitor configuration is artificial in that the progenitor star structure is not self-consistently altered to compensate for the presumed final episodes of mass loss. 
In addition, in some models we rescale the  mass of the progenitor star by simply dividing the density profile everywhere by a constant.
The assumption is that the density profile of our MESA progenitor star provides a reasonable representation of presupernova stars of other masses. In the present context, a simplified approach is not unreasonable in that we will explode the star with a 1D thermal bomb, and the detailed internal structure of the star will be largely washed out by the blastwave. What is most important to the light curve is the structure of the CSM, which in the present case is parameterized in a simplified way that allows us to easily control the physical characteristics.  Future studies using more realistic CSM structures and progenitors are clearly warranted.

\subsection{Hydrodynamical Explosion Simulations}

For modeling the explosion of the star, we use a 1D staggered moving-mesh hydrodynamical code and a gamma-law equation of state with $\gamma = 4/3$, as the SN shock is radiation-pressure dominated. We do not compute the complex mechanism of the explosion itself but instead deposit a
chosen amount of thermal energy $E_{\rm exp}$ at the center of the stellar model to create a thermal bomb. We evolve the explosion until the ejecta profile is roughly homologous, i.e. $r \sim v t$ for all zones. This method has the advantage of speed but is limited to cases in which the CSM radius is small enough that radiative diffusion is not important before homology is reached. 

In the hydrodynamical calculation, some inner zones may remain bound and fall back toward the remnant.  In order to capture this, we use the following criteria to determine if the innermost zone should be ``accreted" and removed from the calculation: 1) the zone has negative velocity; and 2) the gravitational potential energy of the zone exceeds the kinetic and thermal energy of the zone combined by a factor of $1 + \epsilon$, where we typically take $\epsilon$ to be $\sim 0.2$. 
Sometimes an innermost zone will also be removed if its density is some factor $\eta$ larger than the density of the next zone, where $\eta$ is typically $\sim 100$. The density criterion is used because sometimes a zone that is considered unbound by the prior criteria will nevertheless remain spatially small, which imposes a very small time step on the calculation without significantly affecting the results.

\subsection{Radiative Transfer Calculations}

Once our exploded profiles are close to homology, we map the final ejecta structure into SEDONA, a time-dependent Monte Carlo radiation transport code that takes into account the composition, density, and temperature-dependent opacities \citep{kasen06}. We run the code with the assumption of local thermodynamic equilibrium (LTE), which should be reasonable  
for approximating the  phases of the light curve after which interaction with the CSM has taken place, but before the ejecta have become optically thin.

For the models in which we  include $\nifs$ in the ejecta, we  assume the nickel mass fraction $X_{\rm ni}$ profile follows
\begin{equation}
X_{\rm ni} = \frac{1}{2}\biggl( \tanh \biggl[\frac{-(r - r_{\rm ni} )}{s \,dr}\biggr] + 1\biggr)\,\,,
\label{eq:ni_dist}
\end{equation}
where $dr$ is the width of each zone. This equation essentially produces a smoothed step function where $s$ controls the amount of smoothing and the quantity $r_{\rm ni}$ is the shift required, given $s$ to make the total mass of nickel present match a user-specified $M_{\rm ni}$. In this paper, every SEDONA run has the same number of equally spaced radial zones ($N=200$), so $s\,dr$ represents the spatial extent of the smearing and is a fraction of the radial extent of the ejecta controlled by $s$.

\section{Results } \label{s:results}

\subsection{Dynamics of Interaction \label{s:hydro}}

\begin{figure}
\begin{center}
\includegraphics[width=3.5in]{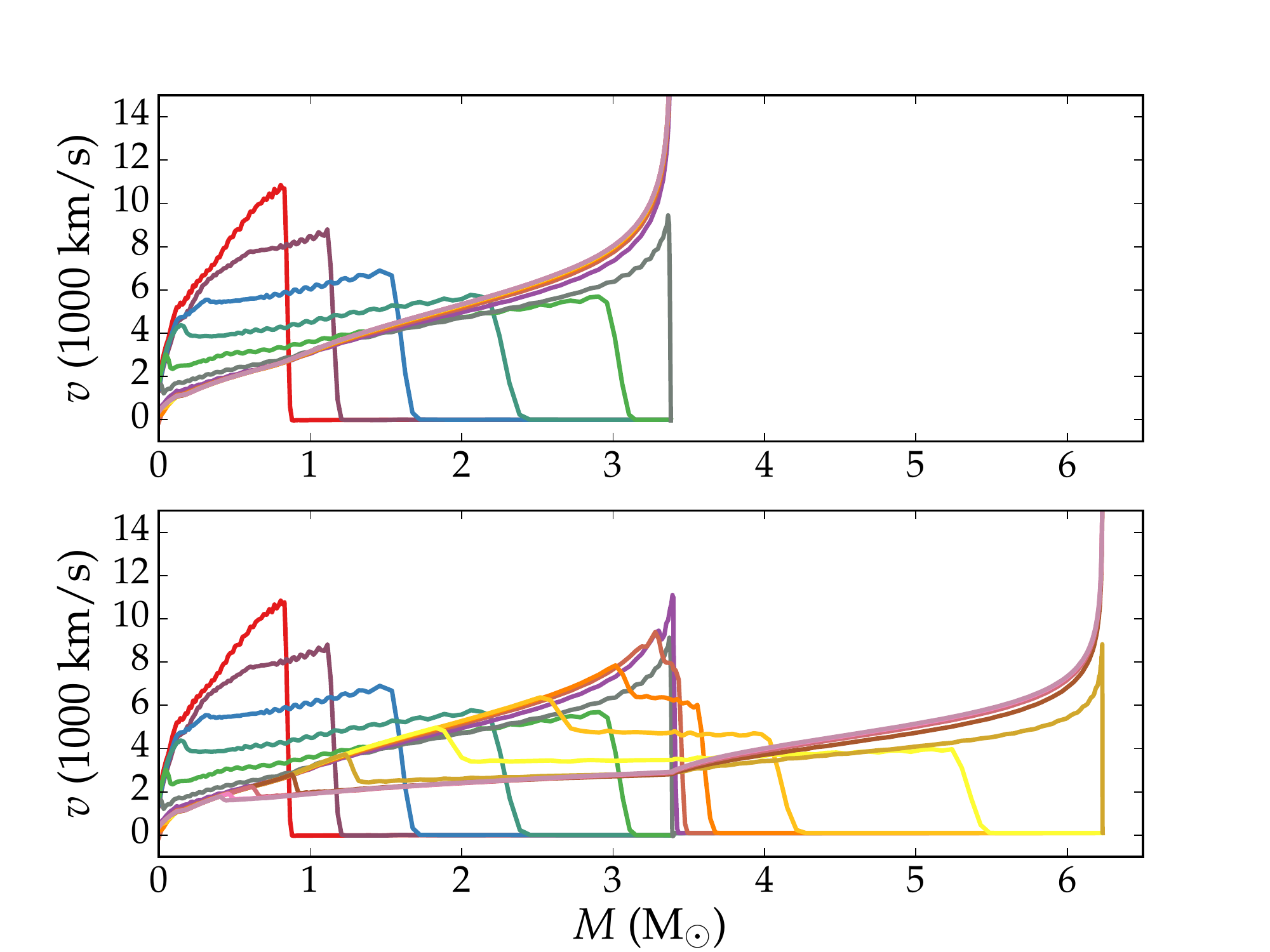}
\end{center}
\caption{Velocity profiles at various times for two hydrodynamical calculations. Each profile corresponds to roughly a doubling in time, i.e. $\sim2~{\rm s}$, $\sim4~{\rm s}$, $\sim8~{\rm s}$, and so forth. Top panel: explosion of a  5 $\msun$ progenitor star ($\sim 3.4~\msun$ once the iron core is removed) with no CSM added. Bottom panel: explosion of the same star with a 3 $\msun$ CSM. The addition of the CSM slows down the forward shock, producing a reverse shock moving toward the center. \label{f:vel_profiles}}
\end{figure}

\begin{figure}
\begin{center}
\includegraphics[width=3.5in]{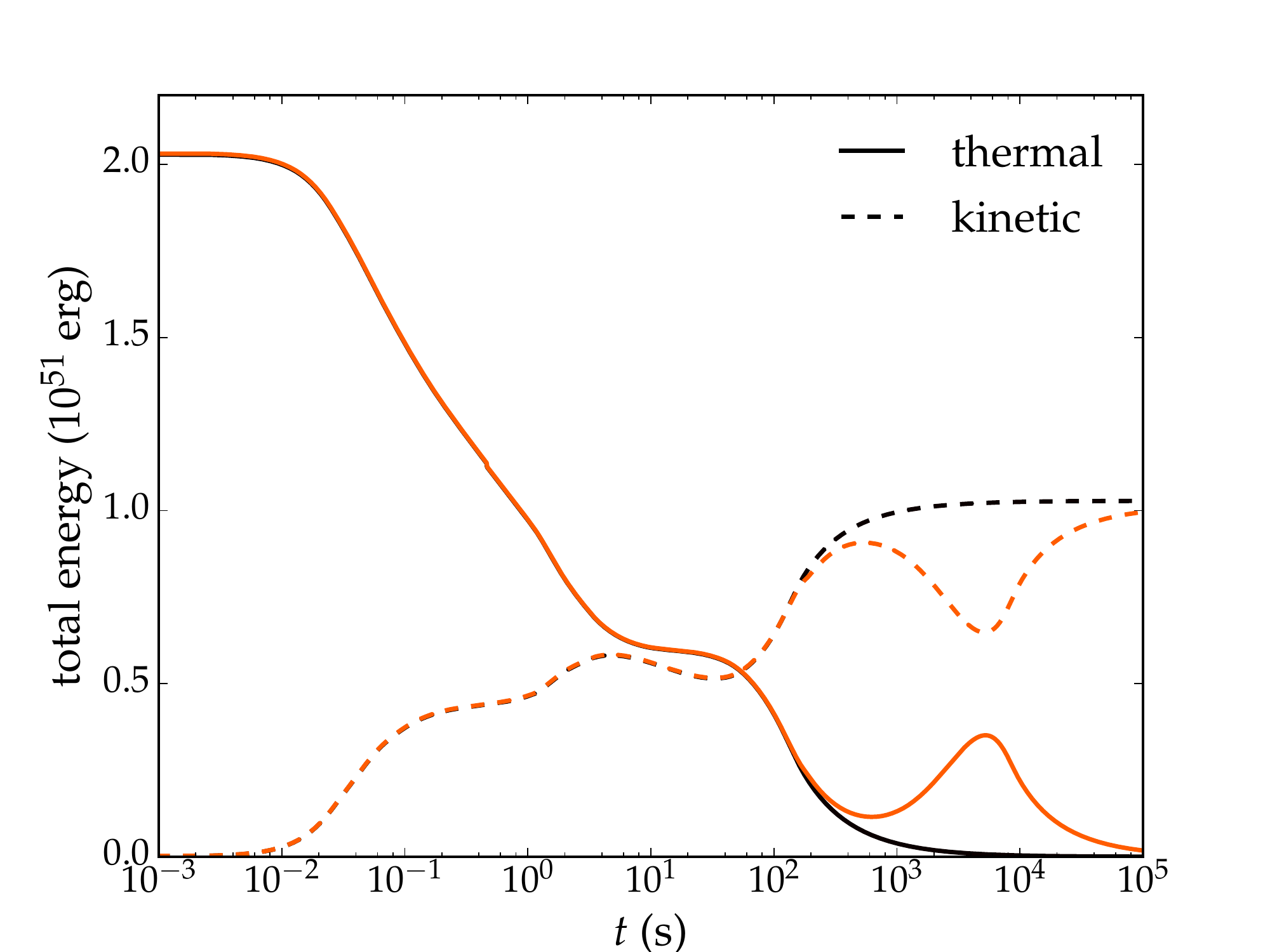}
\end{center}
\caption{Evolution of the total kinetic and thermal energy in the explosion of a $5~\msun$  star with $3~\msun$ of CSM (red lines). For comparison, a model with 
no CSM is also shown (black lines)
A central thermal bomb is input to give an initial thermal energy just above $2~{\rm B}$, resulting in a final kinetic energy of $1~{\rm B}$ once the gravitational potential has been overcome. 
At the earliest times ($t \lesssim 10^2$~s),  thermal energy is converted to kinetic energy as the star explodes. The interaction with the CSM begins at times
$t \gtrsim 10^2$~s and converts  kinetic energy back into thermal energy.  At a time near $10^4$~s, the forward shock breaks out of the CSM. Thereafter the thermal energy declines, closely following the  $t^{-1}$ scaling of adiabatic homologous expansion. \label{f:E_Mbig}}
\end{figure}

\begin{figure*}
\begin{center}
\includegraphics[width=6in]{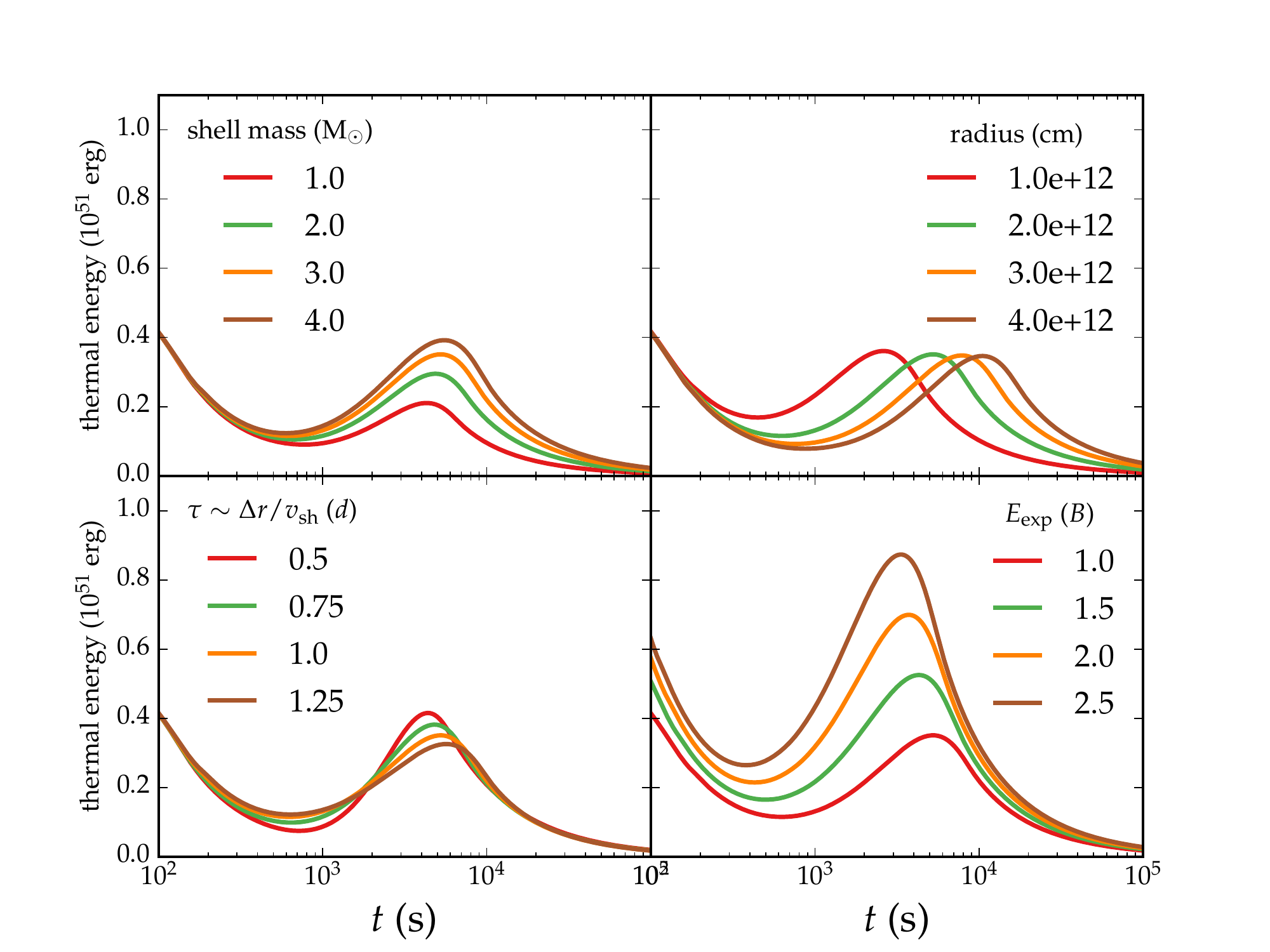}
\end{center}
\caption{Thermal energy evolution for models with different physical parameters. The panels show the effect of varying the CSM mass (top left), CSM radius (top right), CSM thickness (bottom left, note both $r_{\rm mid}$ and $\tau$ are varied proportionally to one another to produce self-similar solutions), and  the explosion energy (bottom right). \label{f:E_fourpanel}}
\end{figure*}

\begin{figure*}
\begin{center}
\includegraphics[width=6in]{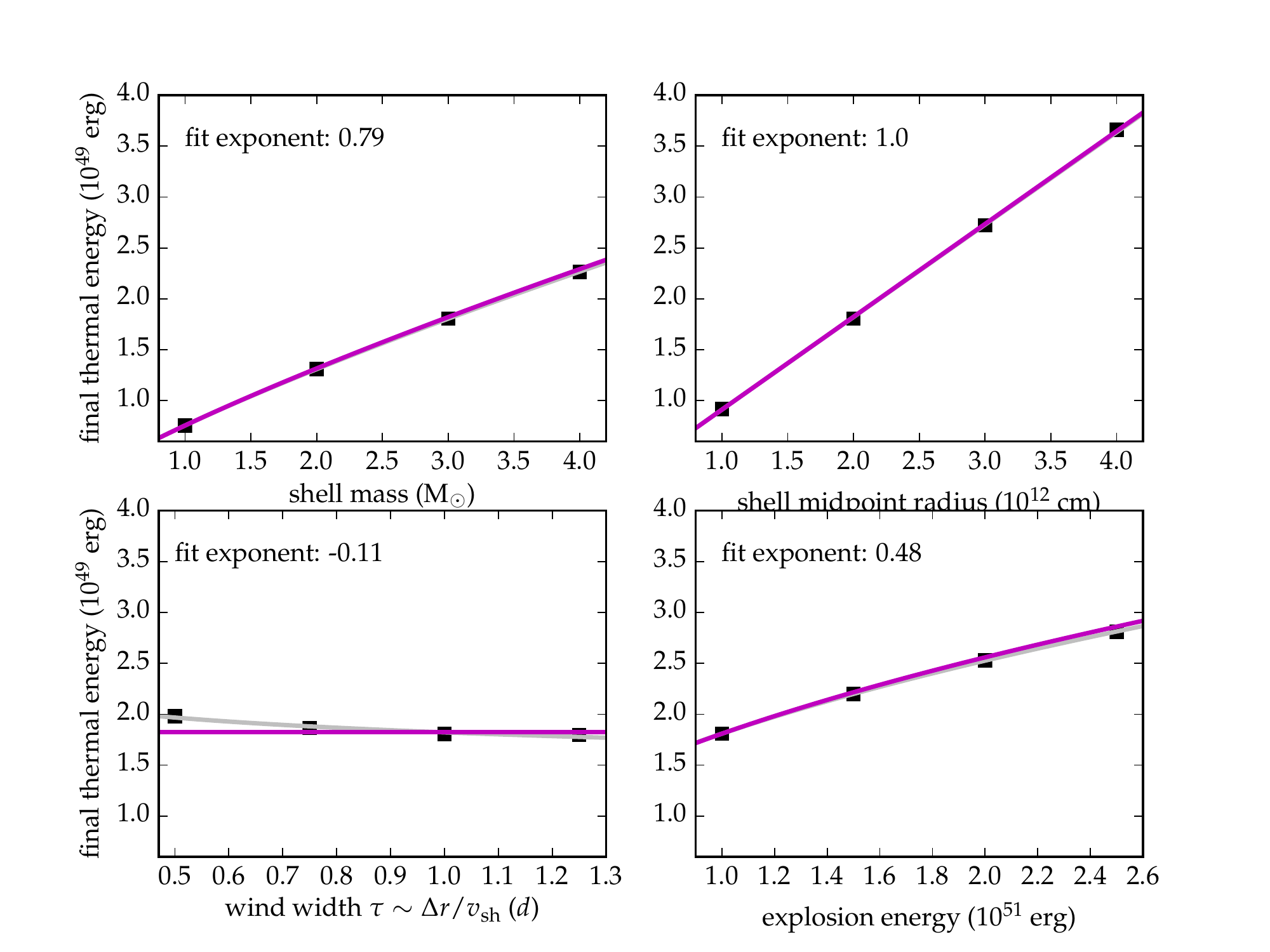}
\end{center}
\caption{Final  thermal energy at $t_{\rm end} = 10^5~{\rm s}$ for each simulation presented in Figure \ref{f:E_fourpanel}. The power-law fits to our numerical data are listed in the figure, and solid gray lines show the fits to the data. Solid magenta lines show our analytical power laws for comparison. The fitted exponents correspond well to our analytical scalings in Equation~\ref{eq:Etht} of \S \ref{s:analytics}. \label{f:fintherm_vary}}
\end{figure*}

\begin{figure}
\begin{center}
\includegraphics[width=3.5in]{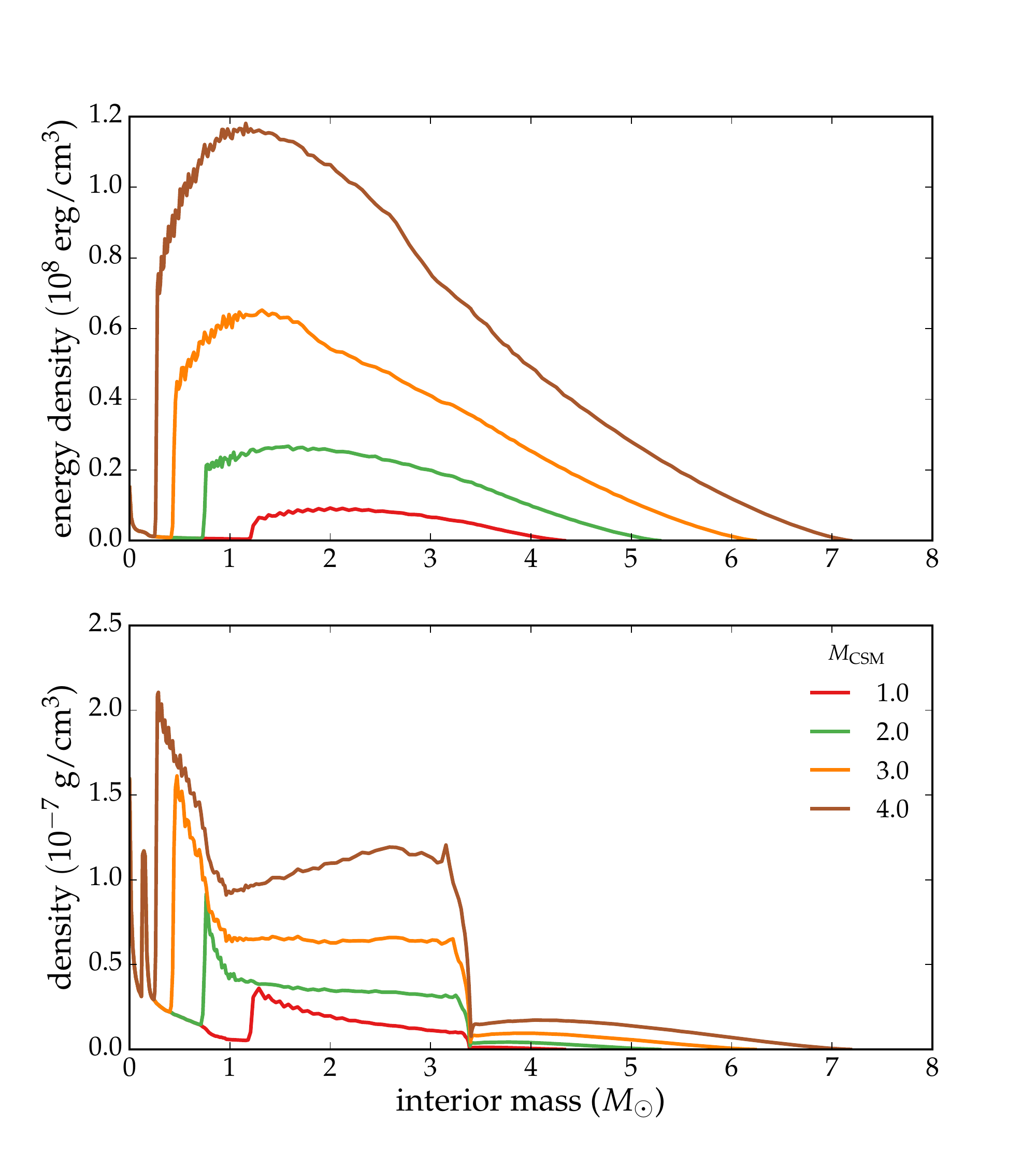}
\end{center}
\caption{Final density and energy density profiles for the explosion of a $5~\msun$ star with different CSM masses. Most of the thermal energy is contained between the reverse shock and
the star/CSM contact discontinuity.  The thermal energy is greater for models with larger CSM masses, and both the density and energy density are concentrated farther inward in mass coordinate. \label{f:prof_vary_M}}
\end{figure}

\begin{figure}
\begin{center}
\includegraphics[width=3.5in]{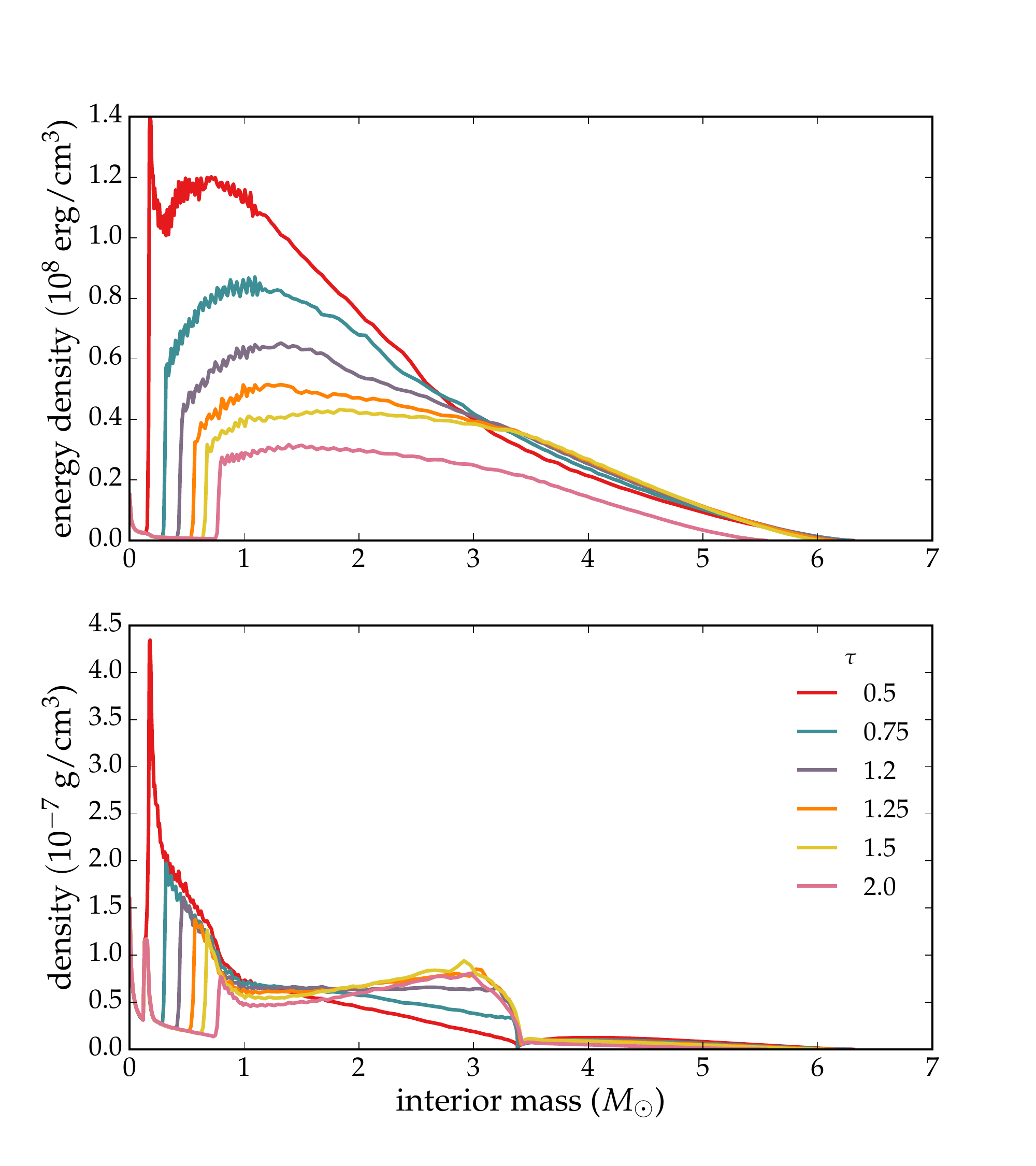}
\end{center}
\caption{Same as Figure \ref{f:prof_vary_M} but for models varying the $\tau$ parameter that sets the CSM thickness. While the CSM thickness does not greatly affect the total thermal energy, it does affect the final distribution of the thermal energy and the location of the reverse shock. \label{f:prof_vary_tau}}
\end{figure}

We present here a study of hydrodynamical simulations of the explosion of the described progenitor star plus CSM configuration. Figure \ref{f:vel_profiles} compares the velocity evolution of a model with no CSM to one with a $3~\msun$ CSM shell. In both models, a strong shock initially propagates outward through the star, reaching the surface (at mass coordinate $3.4~\msun$) at a time $t \approx 10^2$ s. In the model with no CSM, the shock breaks out and accelerates the surface layers of the star to high velocity. In the model with a CSM shell, the interaction produces a reverse shock and a forward shock, the latter of which breaks out of the CSM shell some time later ($t \approx 10^4$ s). The reverse shock weakens after the forward shock breakout due to the pressure release and stalls before reaching the ejecta center.

Figure~\ref{f:E_Mbig} shows the temporal exchange of kinetic and thermal energy in a model with a total kinetic energy at infinity of 1~B.  The thermal energy declines over the intial $\sim 300$ seconds as the shock travels through the star, overcoming the gravitational binding energy and imparting kinetic energy to the stellar material.  In the absence of a CSM shell,  Figure~\ref{f:E_Mbig} shows that the thermal energy  continues to decline to late times due to  expansion loss. In the presence of a CSM shell, however, the outer layers of stellar ejecta impact the shell at $\sim 300$~s and shocks begin to convert kinetic energy back into thermal energy again. The thermal energy content peaks around $5\times 10^3$ seconds, which occurs shortly before the breakout of the forward shock from the CSM. Thereafter, the thermal energy declines again as $1/t$, as expected from $p\,dV$ loses.

\subsubsection{Parameter Study \label{s:param}}

Figure~\ref{f:E_fourpanel}  shows how the thermal energy evolution depends on the ejecta and CSM parameters.  The end result is quantified further in 
Figure~\ref{f:fintherm_vary}, which shows the thermal energy content $E_{\rm th}(t_{\rm end})$ found at a final reference time $t_{\rm end} = 10^5~{\rm s}$. 
The general trends noted are: 1) $E_{\rm th}(t_{\rm end})$ increases with explosion energy, due to the larger available energy budget; 2)  $E_{\rm th}(t_{\rm end})$ increases with
shell mass, due to a larger deceleration and hence thermalization of the ejecta kinetic energy; 3) $E_{\rm th}(t_{\rm end})$ increases with shell radius, as a later onset of interaction leads to less expansion losses by $t_{\rm end}$. Figure~\ref{f:fintherm_vary} demonstrates that the scaling  with these three parameters closely follow the analytic scalings of \S~\ref{s:analytics}. The analytics  did not take into account the shell width, and  Figure~\ref{f:E_fourpanel} shows that it is has a relatively small impact on the final thermal energy content.

The radial density and energy density distributions  of our exploded models at $t_{\rm end}$ are shown in Figures \ref{f:prof_vary_M} and \ref{f:prof_vary_tau}. The density profiles show two  sharp features,  one at the location where the inward propagating reverse shock stalled, and one at the location of the contact discontinuity between the star and CSM. The  energy density has a smoother radial distribution.
 Figure \ref{f:prof_vary_tau} shows that, even though the shell width does not impact the total thermal energy content, it does affect the radial distribution, with more extended shells leading to more central concentration of mass and energy. This will have some effect of the shape of the resulting light curve.

\subsubsection{Fallback }\label{s:fallback}

For models with strong interaction, the
reverse shock may reach the center of the ejecta and induce fallback onto the remnant
\citep[e.g.,][]{chevalier89b}. Alternatively, low explosion energies could also allow larger amounts of mass to remain bound to the remnant.
It is interesting to speculate whether this fallback could provide a mechanism to explain the apparently low $\nifs$ masses inferred for some RFSNe, as  $\nifs$ is synthesized in the innermost layers of the star.
Following previous work on SN fallback \citep[see e.g.][]{mcfadyen01, zhang08}, we explore here the amount of material which may remain bound to the central remnant following the explosion.

\begin{figure*}
\begin{center}
\includegraphics[width=7in]{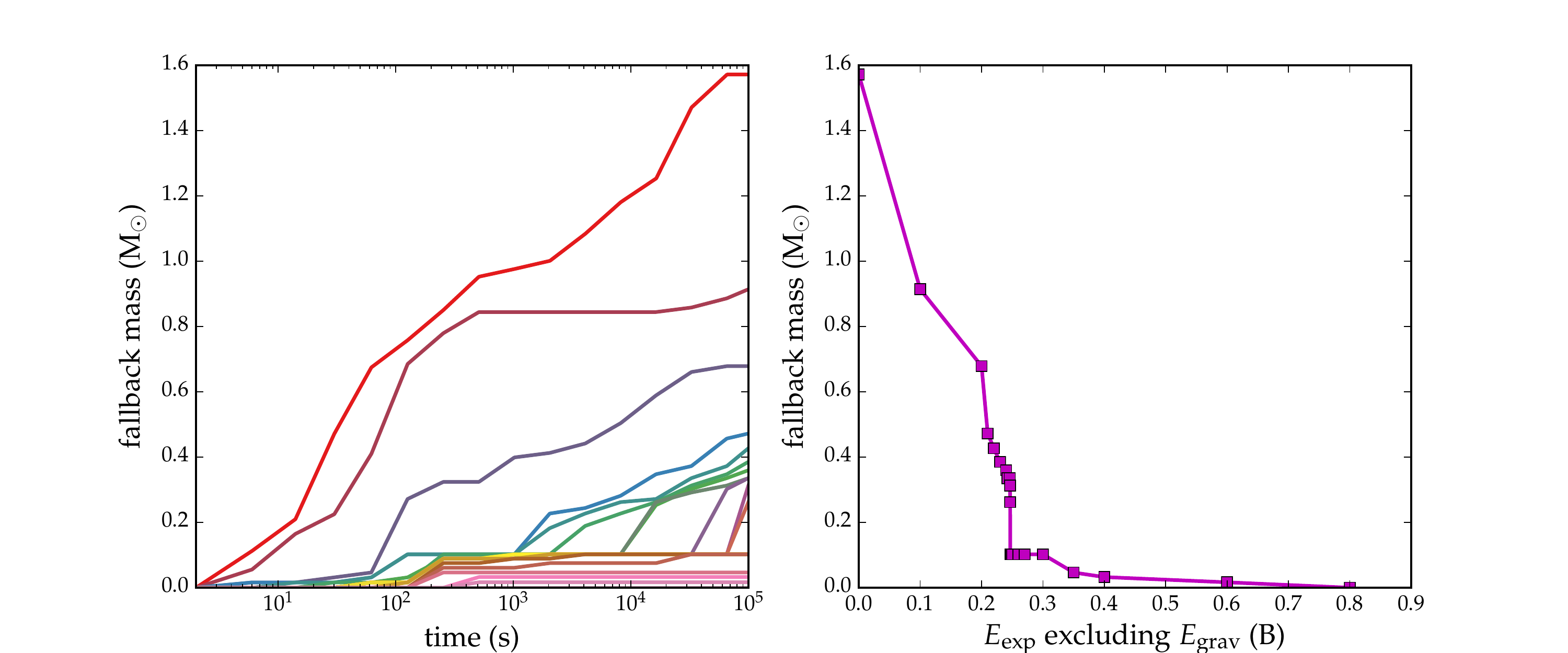}
\end{center}
\caption{Amount of fallback in the explosion of  a 5~$\msun$ star with $3~{\rm M}_\odot$ of CSM. Left: Cumulative fallback mass over time for models with various explosion energies. Right: Final amount of fallback as a function of explosion energy. Here explosion energy refers to the final kinetic energy of the ejecta at infinity. For lower energies ($E < 0.5$~B) the fallback mass can be
significant ($\gtrsim 0.05~\msun$) and may influence the mass of radioactive $\nifs$ ejected. \label{f:fallback}}
\end{figure*}

Figure~\ref{f:fallback} shows the amount of fallback for models with $3~\msun$ of CSM and various explosion energies. For models with $E = 1$~B the fallback mass is small ($\lesssim 0.01~\msun$).
This is because the reverse shock stalls before reaching the ejecta center. A CSM mass of $\mcsm \gtrsim M_{\rm ej}$ is needed for the reverse shock
to approach the center in a $E = 1$~B explosion (see Figure~\ref{f:prof_vary_M}).

For low explosion energies ($E \lesssim 0.3-0.5$~B) and $\mcsm \approx M_{\rm ej}$ the fallback mass may 
be significant, $\gtrsim 0.05~M_\odot$. This is comparable to
the typical mass of $\nifs$ inferred to be ejected in  core collapse SNe. Since  $\nifs$ is synthesized in the densest, innermost regions, such strong fallback could significantly reduce or eliminate entirely the radioactivity available to contribute to the light curve.

The results in Figure~\ref{f:fallback} are only suggestive, as the actual amount of fallback will depend on the details of the progenitor structure and explosion mechanism. 
Whether fallback is relevant for RFSNe is unclear. Given the scalings of Figure~\ref{f:fintherm_vary}, a low explosion energy will lead to a dim light curve unless the progenitor star radius is very large. Alternatively, if the explosion energy is typical ($E \approx 1$~B), the 
CSM mass likely needs to exceed that of the ejecta. Even in cases where the fallback mass is significant, multi-dimensional effects could mix synthesized $\nifs$ out
to larger radii, allowing some radioactive material to be ejected. More detailed simulations are needed to evaluate the importance of fallback in RFSNe.

\subsection{Light Curves \label{s:lc}}

\subsubsection{Nickel-Free Light Curves \label{s:nifree}}

Having run hydrodynamical simulations of the ejecta/CSM interaction, we post-process the results with radiation transport calculations in SEDONA.
Table~\ref{t:many_lc} gives the parameters of the models considered, along with our calculated rise time, decline time, and peak brightness. 
 Figure \ref{f:lc_10x}  shows a specific example light curve compared to data from \snx. While the parameters ($\msh = 3.0~\msun$, $R_{\rm mid} = 2 \times 10^{12}~{\rm cm}$, $\tau = 1~{\rm day}$, $E_{\rm exp} = 3~{\rm B}$)
were not finely tuned to fit this particular object, the model reproduces the bulk properties of this supernova rather well.

\begin{figure*}
\begin{center}
\includegraphics[width=3in]{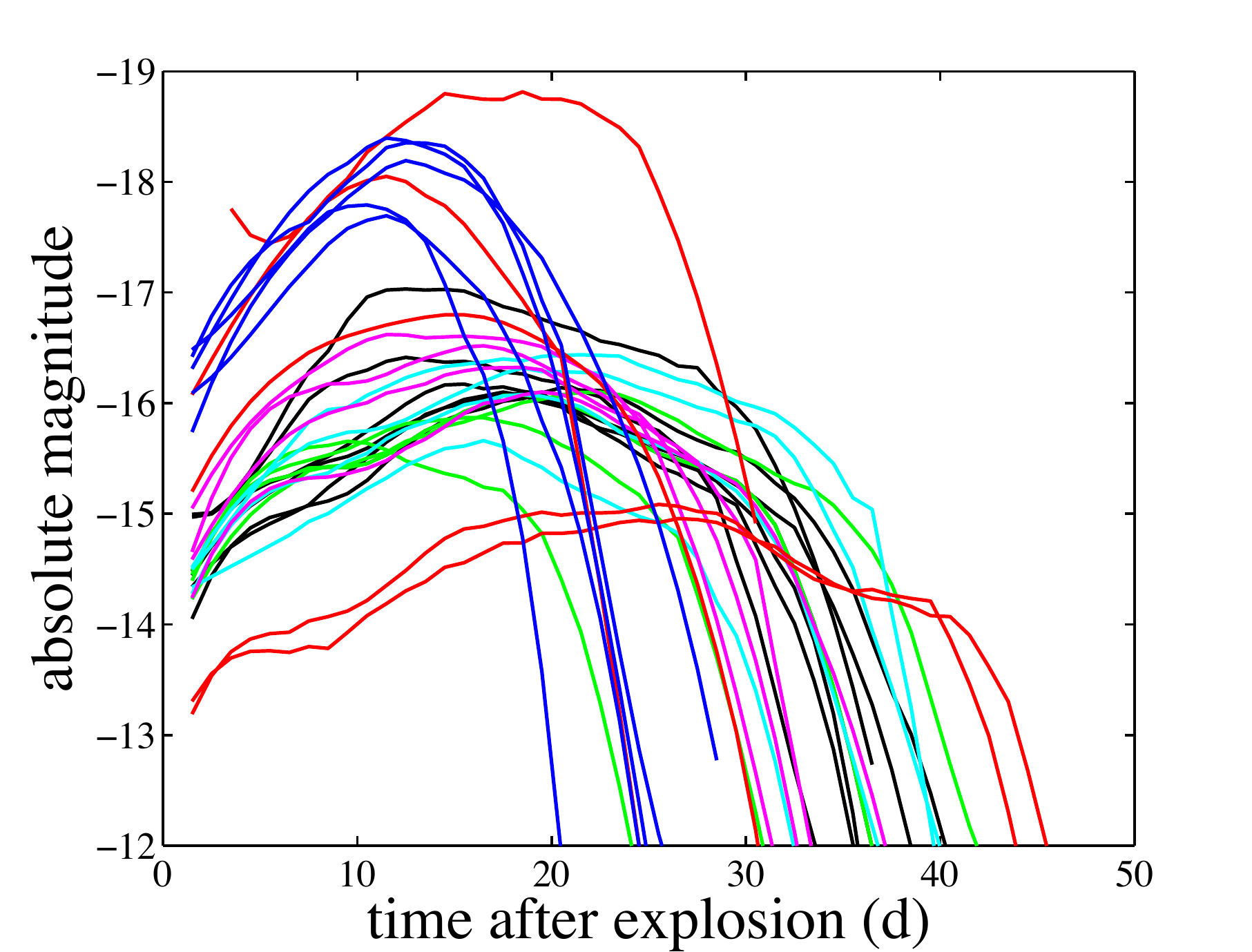}
\includegraphics[width=3in]{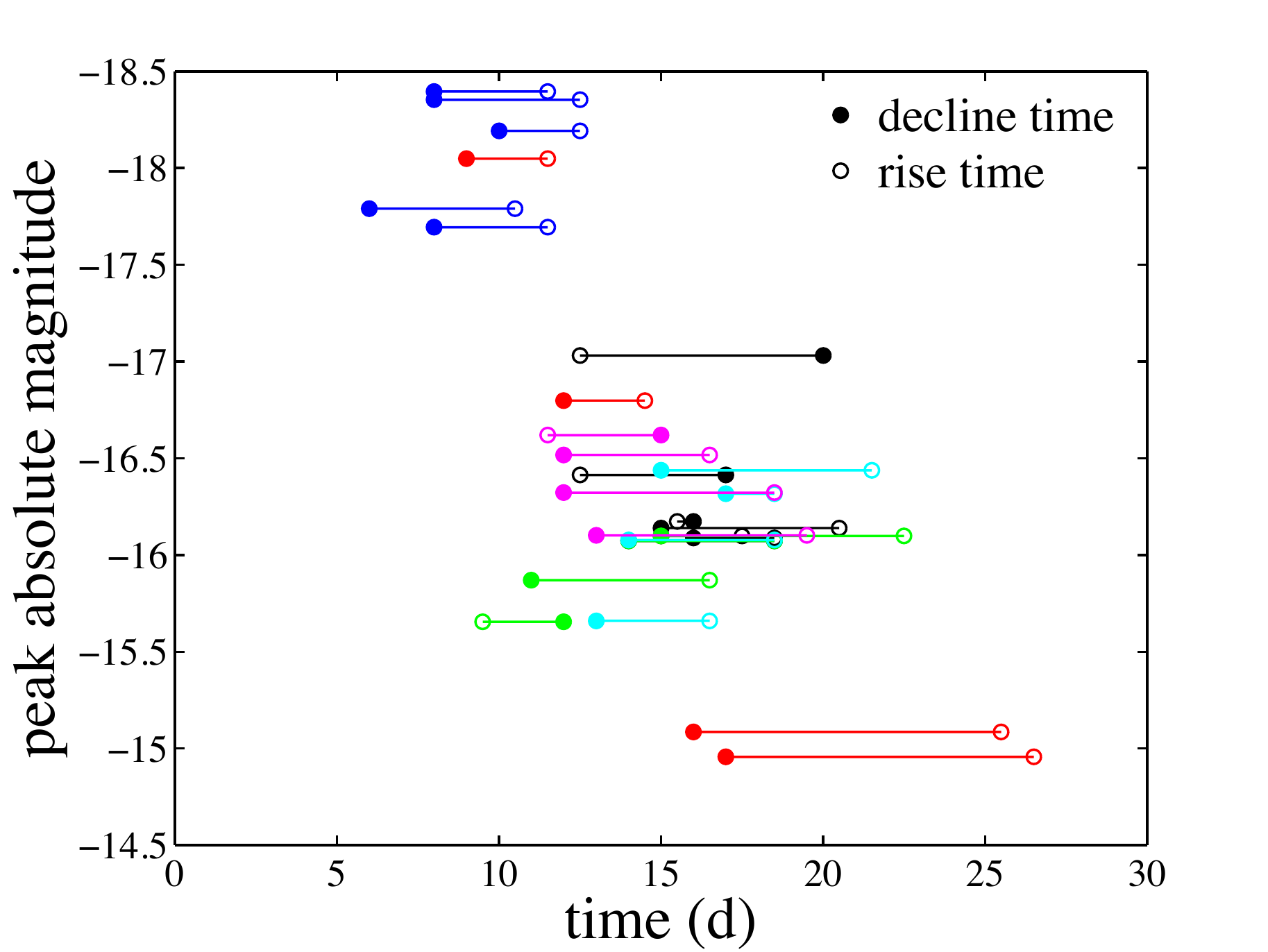}
\end{center}
\caption{Calculated $r$-band optical data for many of the hydrodynamical models from Section \ref{s:hydro}. Left: Light curves including parameter variation in radius, explosion energy, shell mass, and $\tau$. This plot also includes more extreme runs with large energy $E_{\rm exp} = 6~B$ and fallback models with $E_{\rm exp} = 0.22,~0.25~B$. Light curves have been run with low photon counts for speed and then smoothed using Savitzsky-Golay filtering. Right: Peak magnitude and timescale plots for these light curves. To the left of the plot is the rise time ($t_{\rm peak} - t_0$). To the right are decline times determined by how long it takes for the $r$-band light curve to decline from peak by two magnitudes. The parameters and bulk properties of the runs plotted here are shown in Table 
\ref{t:many_lc}.
\label{f:many_lc}}
\end{figure*}

We show in Figure \ref{f:many_lc}  the variety  of $r$-band light curves and bulk properties
(peak brightness, rise time, and decline time) for our parameter survey of different CSM structures and explosion energies. 
Similar to the observed diversity in RFSNe shown by \citet{drout14}, the model light curves 
display generally short durations but span a wide range in brightness. For the parameter range chosen,
 most of our models  occupy the lower-luminosity ($M_r > -17$) region. However, models with higher explosion energies $(E > 1$~B) or larger radii $R_{\rm csm} \gtrsim 10^{14}~$cm, and lower ejected masses ($M \lesssim 2 M_\odot$) begin to approach the luminosity and rapid timescales of the brightest RFSNe.

To explore the effect of ejecta mass in a parameterized way, we have also included in our sample a model for which the stellar density profile has been reduced by a factor of 3 and exploded into a $1~{\rm M}_\odot$ shell with 3~B. The resulting light curve is very similar to that of the original mass star exploded into a $1~{\rm M}_\odot$ shell with 6 B, suggesting that the structure of the star itself is not particularly important to the shape of the light curve but rather that the $E/M$ ratio and CSM structure primarily determine the gross properties of the observed supernova.

While the properties of the models in our parameter survey resemble those of many observed RFSNe, the models do not well fit the light curves of some higher-luminosity events. As shown in Figure \ref{f:many_lc}, while we can attain the necessary peak luminosities and timescales for \snbj and \snu, the shapes of the light curves are different; in particular, it is difficult to obtain a short enough rise time to match the observations. This indicates
that the fastest rising events may not be explained by post-shock cooling. A fast ($\sim$days) rise of the light curve may be possible as a result of
shock breakout  in  dense CSM \citep{chevalier11}.  It is also possible that
in some events, significant CSM interaction is ongoing throughout the light curve. The narrow He lines seen in \snu \citep{shivvers16} certainly suggest that there is ongoing conversion of kinetic energy to thermal energy, well past the supernova peak. Capturing these properties would require the use of radiation-hydrodynamics calculations (rather than treating the hydrodynamics and radiation transport separately in sequence).

\begin{figure*}
\begin{center}
\includegraphics[width=6in]{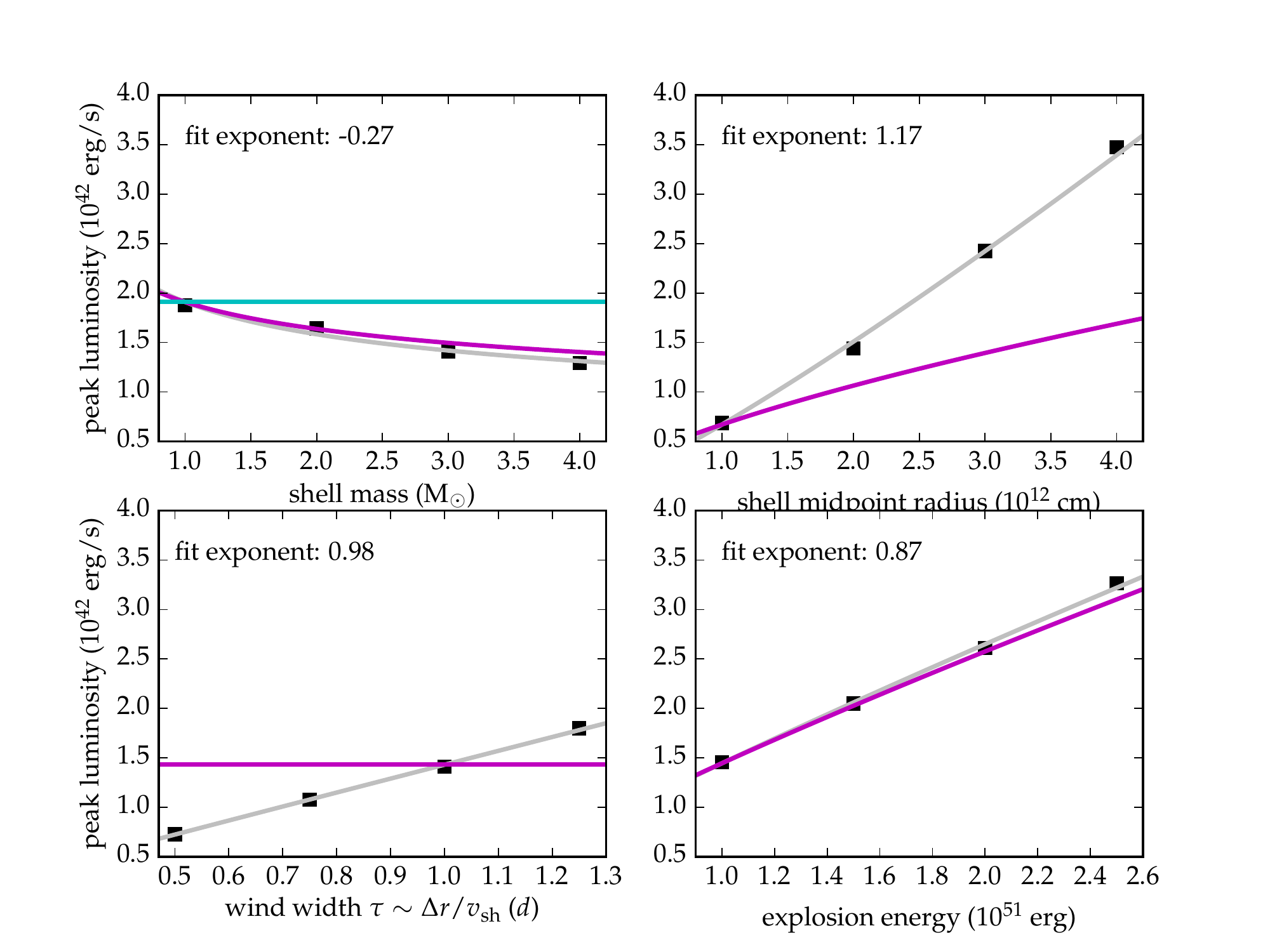}
\end{center}
\caption{Peak luminosities for the parameter study shown in Figure \ref{f:fintherm_vary}. The power-law fits to our numerical data are listed in the figure, and solid gray lines show the fits to the data. Solid magenta lines show our analytical power laws from Equation~\ref{eq:L_sn_b} of \S \ref{s:analytics} using $n=6$. The cyan line in the first panel represents the same but using $n=8$ for the mass variation. Note that there is a stronger dependence of $L_{\rm sn}$ on both $\tau$ and $R_{\rm CSM}$, which we tentatively attribute to the different distribution of energy for different CSM structures, as shown in Figure \ref{f:prof_vary_tau}. \label{f:Lpeak_vary}}
\end{figure*}

\begin{figure*}
\begin{center}
\includegraphics[width=6in]{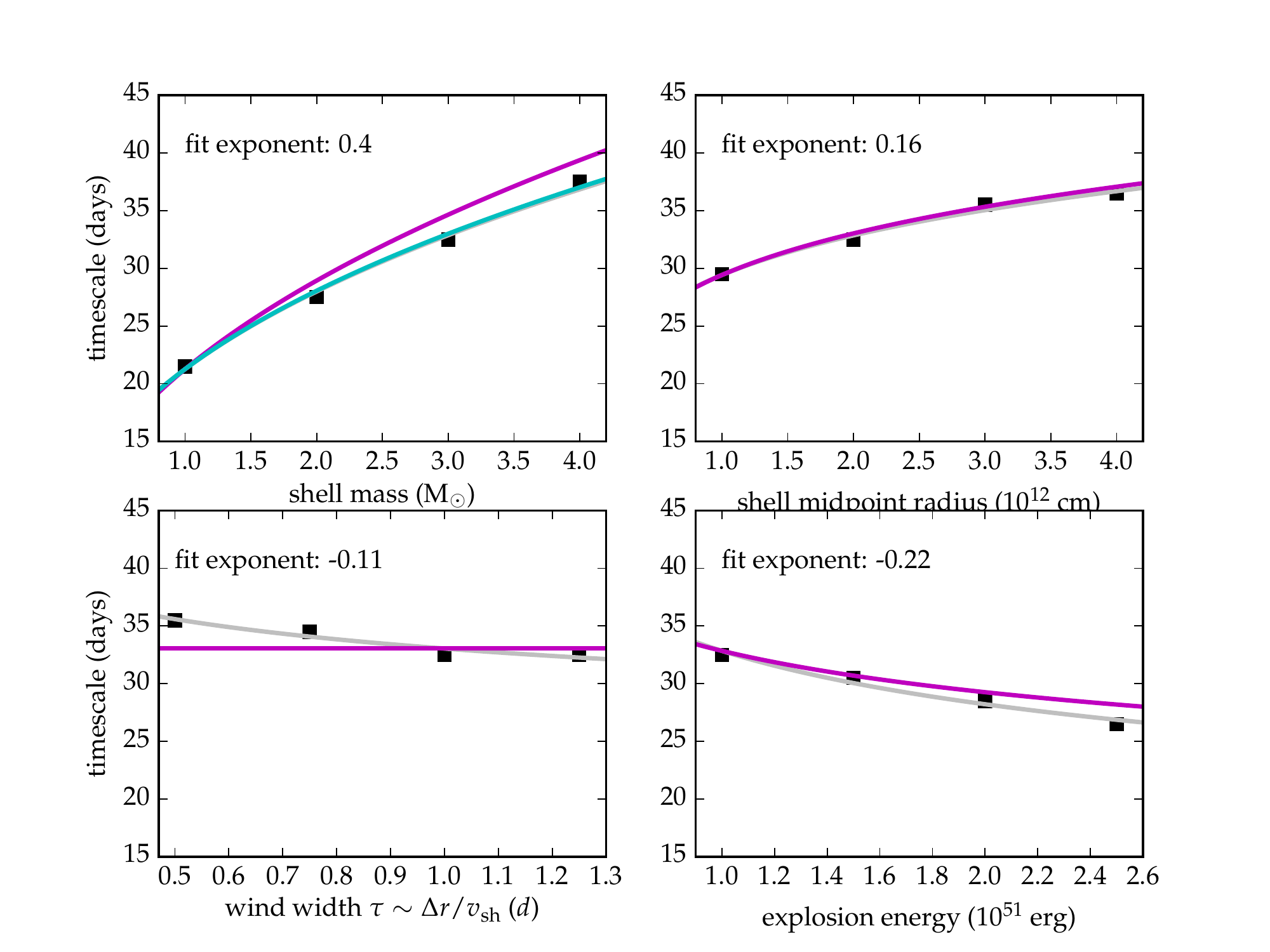}
\end{center}
\caption{Same as Figure \ref{f:Lpeak_vary} but for timescales $t_{\rm sn} = t_{\rm rise} + t_{\rm decline}$.  Again, gray lines show our power-law fits to the data, while magenta lines show analytic results from Equation~\ref{eq:t_sn_b} of \S \ref{s:analytics}. As in Figure Figure \ref{f:Lpeak_vary}, the magenta line in the first panel uses $n=6$, and the cyan line uses $n=8$. \label{f:timescale_vary}}
\end{figure*}

Figures \ref{f:Lpeak_vary} and \ref{f:timescale_vary} show numerical versus analytical results for the same series as presented in Figure \ref{f:fintherm_vary}. While our analytical estimates for the total available energy were quite accurate, the light curves are somewhat more complex. Because $t_{\rm sn}$ and $L_{\rm sn}$ depend on both the sum and ratio of $M_{\rm CSM}$ and $M_{\rm ej}$ in Equations \ref{eq:t_sn} and \ref{eq:L_sn}, they do not lend themselves to simple power laws because of the $M_{\rm diff}$ factor. As we showed subsequently in \S \ref{s:analytics}, there are some assumptions that can be used to simplify these expressions. In these figures, we have plotted the examples using $t_{\rm sn} \propto M_{\rm CSM}^{\frac{-(n-5)}{6(n-3)} + \frac{1}{2}}$ and $L_\mathrm{sn} \propto M_{\rm CSM}^{\frac{5(n-5)}{6(n-3)} - \frac{1}{2}}$ with $n=6$ and $n=8$ as examples.

We also see that, while our analytics did not consider the effects of varying the shell width $\tau$, $L_{\rm sn}$ shows a nearly linear dependence on this parameter. This may be because a more diffuse shell produces a weaker reverse shock and more evenly distributes thermal energy in the ejecta (see Figure \ref{f:prof_vary_tau}), allowing for a higher and earlier peak. We also see a much larger dependence on radius than expected, possibly in part due to the fact that when increasing the radius we also increased $\tau$ proportionally such that the profile of the ejecta would simply scale.

We also derive scalings from our numerical results, including for $\tau$, which was not included in our analytical predictions. Equations for peak luminosity and timescale based on the fits to our numerical results are:
\begin{equation}
L_{\rm sn} \approx (1.3\times10^{42}~{\rm erg/s}) 
~M_{\rm CSM}^{-0.27}R_0^{1.17}\tau^{0.98}E_{\rm exp}^{0.87}\,\,,
\end{equation}
\begin{equation}
t_{\rm sn} \approx (29~{\rm days}) 
~M_{\rm CSM}^{0.4}R_0^{0.16}\tau^{-0.11}E_{\rm exp}^{-0.22}\,\,.
\end{equation}
The normalizations are obtained by taking the average value from the fits to each parameter variation and then reducing to one significant figure due to the uncertainty.

\subsubsection{Spectra for SN 2010X}

\begin{figure}
\begin{center}
\includegraphics[width=3.5in]{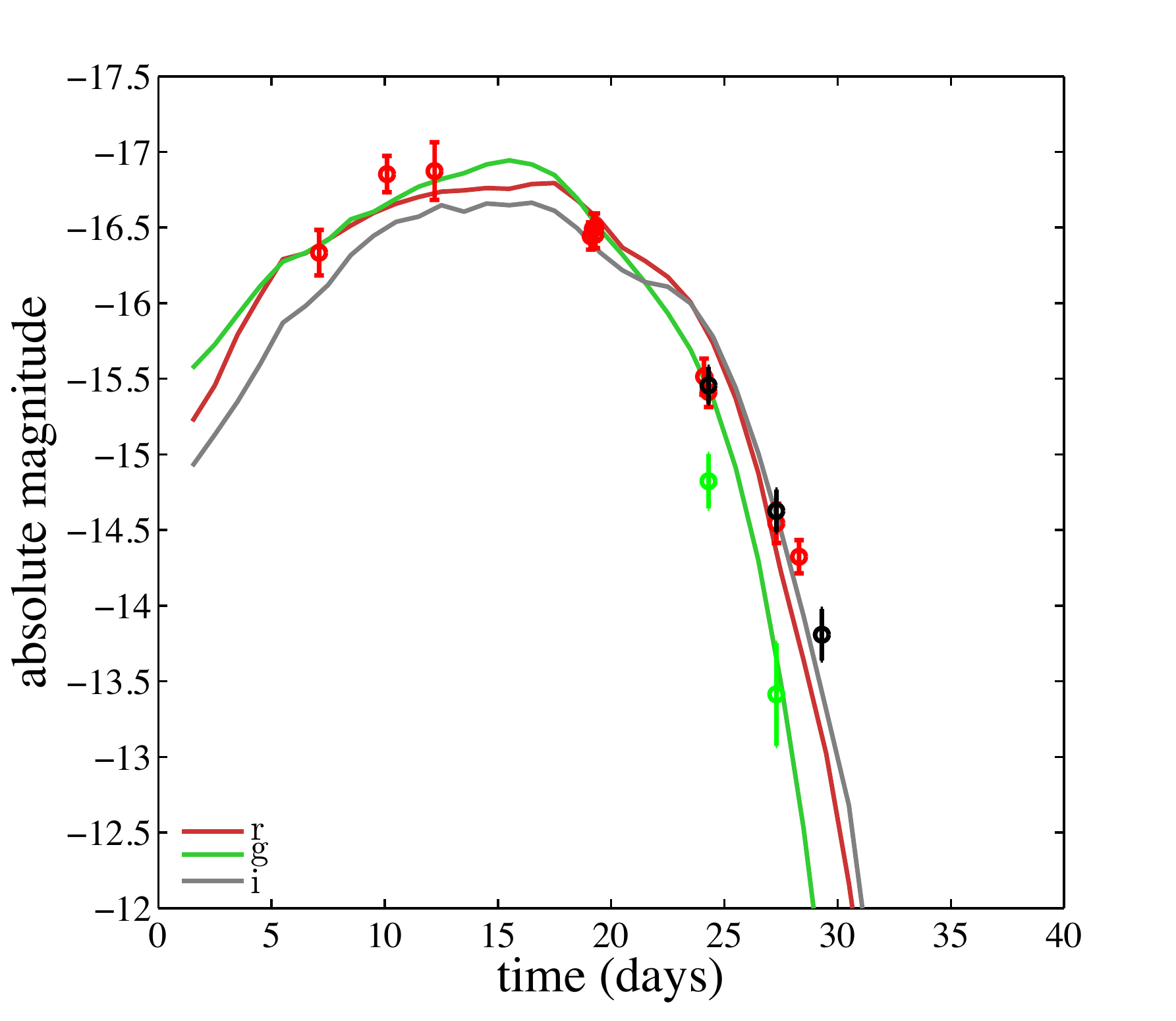}
\end{center}
\caption{Light curve from one run plotted against the light curves for \snx. The parameters used here are $E_{\rm exp} = 3~B$, $\msh = 3~\msun$, $r_{\rm mid} = 2\times10^{12}~{\rm cm}$, and $\tau = 1~{\rm day}$. Because the parameters were not specifically tuned, we do not expect a perfect fit, but this comparison is to demonstrate the viability of the shock cooling model to explain main RFSNe even without extensive model tweaking. We correct the data for Galactic extinction along the line of sight to the host galaxy, NGC 1573A: $A_g = 0.483$; $A_r = 0.334$; $A_i = 0.248$. We do not assume host galaxy extinction. \label{f:lc_10x}}
\end{figure}

\begin{figure*}
\begin{center}
\includegraphics[width=6in]{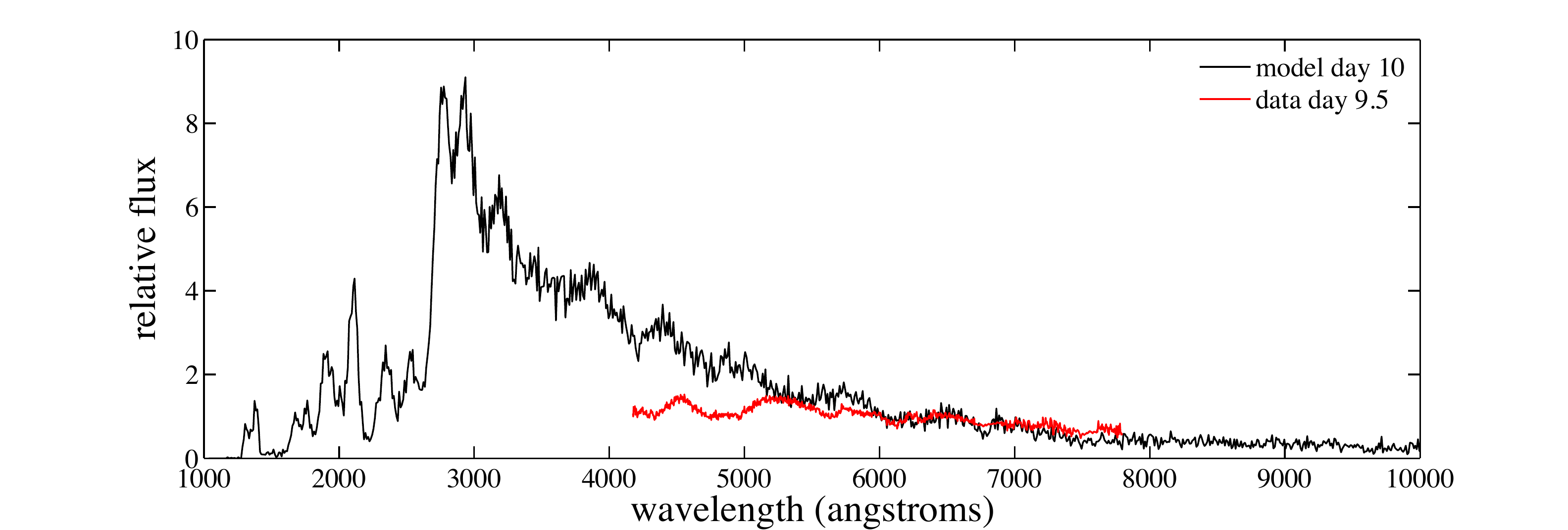}
\includegraphics[width=6in]{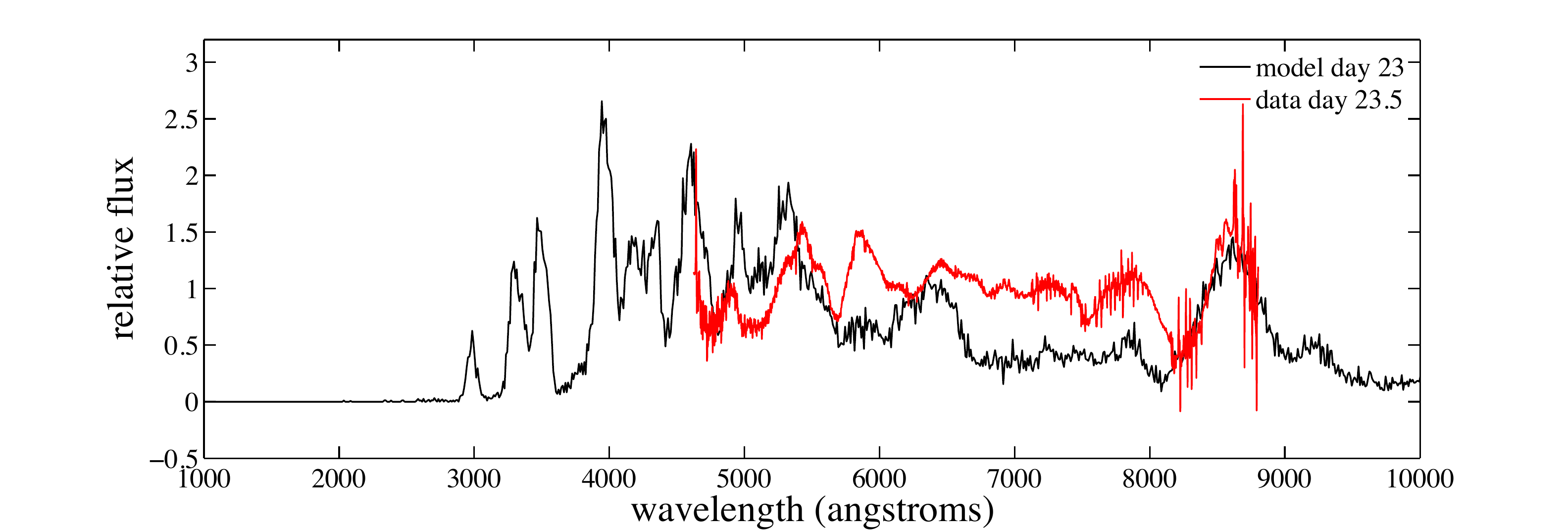}
\end{center}
\caption{Spectra of the same model shown in Figure \ref{f:lc_10x} at days 12 and 23 after explosion (black). We have plotted data from \snx at days 9.5 and 23.5, respectively, for comparison (red), after correcting for redshift and Galactic extinction. The presumed day after explosion for the data is determined by the shift we use in matching the light curve data to our model light curves. Note that many of the same features are reproduced, but the relative strengths can differ for a variety of possible reasons, including variations in composition, temperature, and ejecta structure. Because we have not finely tuned our model to fit this object, we expect it to recover only the bulk properties of the spectra, which is typical of SNe Ibc. Our calculated spectra are also slightly bluer, which could be corrected by assuming some amount of extinction for the host galaxy. \label{f:spec_10x}}
\end{figure*}

While a comprehensive study of the spectroscopic properties of our models is beyond the scope of this work, we show in Figure \ref{f:spec_10x} example spectra of the single \snx model whose light curve is shown in Figure \ref{f:lc_10x}. Figure \ref{f:spec_10x} shows comparisons of our calculated spectra to those obtained by \citet{kasliwal10} at similar days. The observed spectra have been corrected for the redshift of the host galaxy (NGC 1573A at $z = 0.015014$) and de-reddened using Galactic extinction value along the line of sight $A_V = 0.401$ but assuming no host extinction. As can be expected, the results from our model resemble those of a typical SN Ibc, although at early times they are quite blue. They compare fairly well with \snx spectra, showing many of the same features but not always recovering their relative strengths. The calculated spectra are also slightly bluer across the board, which could be due to unaccounted-for host extinction that we have chosen to exclude from our corrections to the data.

\clearpage

\onecolumn

\begin{table*} 
\footnotesize
\caption{ Table of values presented in Figure \ref{f:many_lc}.}
\centering 
\begin{tabular}{c c c c c c c c} 
\hline\hline 
$M_{\rm shell}~({\rm M}_\odot)$ & $\tau~({\rm d})$ & $R_{\rm mid}$ & $E_{\rm exp}$ & $M_{\rm peak}$ & decline time (d) & rise time (d) & color plotted \\ [0.5ex] 
3.0  &  0.5   &  $2 \times 10^{12}$  &   1.0  &  -16.1391   &   15  &   20.5  & black \\
3.0  &  0.75  &  $2 \times 10^{12}$  &   1.0  &  -16.0881   &   16  &   18.5  & black \\
3.0  &  1.0   &  $2 \times 10^{12}$  &   1.0  &  -16.0729   &   14  &   18.5  & black \\
3.0  &  1.25  &  $2 \times 10^{12}$  &   1.0  &  -16.0990   &   15  &   17.5  & black \\
3.0  &  1.4   &  $2 \times 10^{12}$  &   1.0  &  -16.1721   &   16  &   15.5  & black \\
3.0  &  4.0   &  $2 \times 10^{12}$  &   1.0  &  -16.4137   &   17  &   12.5  & black \\
3.0  &  10.0  &  $2 \times 10^{12}$  &   1.0  &  -17.0311   &   20  &   12.5  & black \\
\hline
1.0  &  1.0  &  $2 \times 10^{12}$  &   1.0  &  -15.6547   &   12  &    9.5  & green \\
2.0  &  1.0  &  $2 \times 10^{12}$  &   1.0  &  -15.8692   &   11  &   16.5  & green \\
3.0  &  1.0  &  $2 \times 10^{12}$  &   1.0  &  -16.0729   &   14  &   18.5  & green \\
4.0  &  1.0  &  $2 \times 10^{12}$  &   1.0  &  -16.0987   &   15  &   22.5  & green \\
\hline
3.0  &  0.5  &  $1 \times 10^{12}$  &   1.0  &  -15.6602   &   13  &   16.5  & cyan \\
3.0  &  1.0  &  $2 \times 10^{12}$  &   1.0  &  -16.0774   &   14  &   18.5  & cyan \\
3.0  &  1.5  &  $3 \times 10^{12}$  &   1.0  &  -16.3172   &   17  &   18.5  & cyan \\
3.0  &  2.0  &  $4 \times 10^{12}$  &   1.0  &  -16.4376   &   15  &   21.5  & cyan \\
\hline
3.0  &  1.0  &  $2 \times 10^{12}$  &   1.0  &  -16.1009   &   13  &   19.5  &  magenta \\
3.0  &  1.0  &  $2 \times 10^{12}$  &   1.5  &  -16.3225   &   12  &   18.5  &  magenta \\
3.0  &  1.0  &  $2 \times 10^{12}$  &   2.0  &  -16.5175   &   12  &   16.5  &  magenta \\
3.0  &  1.0  &  $2 \times 10^{12}$  &   2.5  &  -16.6194   &   15  &   11.5  &  magenta \\
\hline
3.0  &  10.0  &  $2 \times 10^{12}$  &   6.0   &  -18.0486   &    9  &   11.5  & red \\
3.0  &  1.0   &  $2 \times 10^{12}$  &   0.22  &  -14.9561   &   17  &   26.5  & red \\
3.0  &  1.0   &  $2 \times 10^{12}$  &   0.25  &  -15.0852   &   16  &   25.5  & red \\
3.0  &  1.0   &  $2 \times 10^{12}$  &   3.0   &  -16.7979   &   12  &   14.5  & red \\
1.0*  &  (wind)   &  $2 \times 10^{14}$  &   3.0   &  -18.8123   &   9  &   18.5  & red \\
\hline
1.0  &  10.0  &  $1 \times 10^{12}$  &   3.0  &  -17.6945   &    8  &   11.5  & blue \\
1.0  &  10.0  &  $1 \times 10^{12}$  &   6.0  &  -17.7910   &    6  &   10.5  & blue \\
1.0  &  10.0  &  $2 \times 10^{12}$  &   3.0  &  -18.1923   &   10  &   12.5  & blue \\
1.0  &  10.0  &  $2 \times 10^{12}$  &   6.0  &  -18.3955   &    8  &   11.5  & blue \\
1.0*  &  10.0  &  $2 \times 10^{13}$  &   3.0  &  -18.3533   &    8  &   12.5  & blue \\

\hline 
\end{tabular} \\
* Stellar model with density profile reduced by a factor of three in order to explore lower ejecta mass. \\
The label (wind) signifies that in this case the CSM density profile goes as $r^{-2}$ and is not modified by the Gaussian.

\label{t:many_lc} 
\end{table*} 
\normalsize

\twocolumn

\subsubsection{Double-Peaked Light Curves \label{s:dbl_pk}}

\begin{figure}
\begin{center}
\includegraphics[width=3.5in]{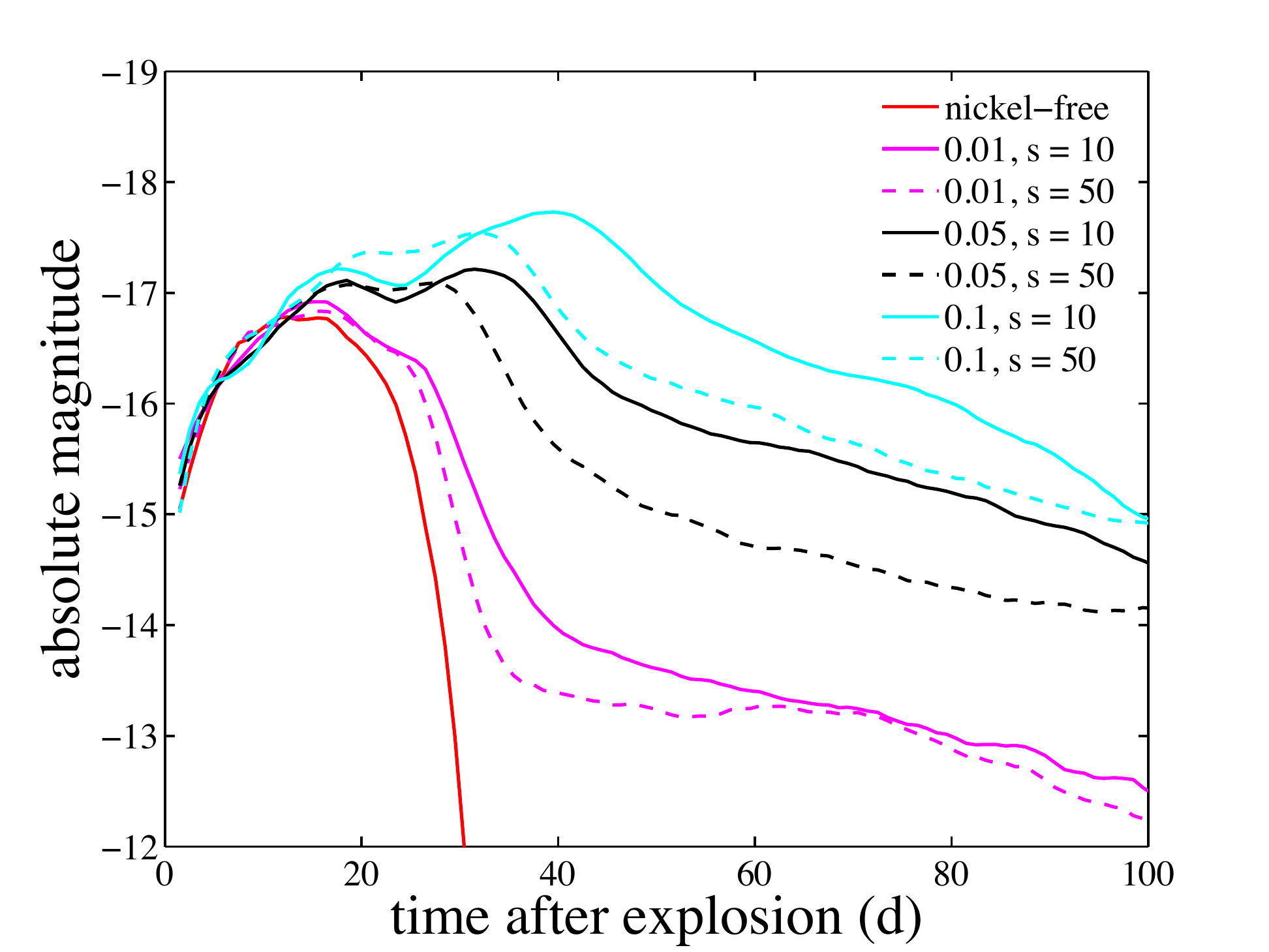}
\end{center}
\caption{Model light curves obtained by adding  $\nifs$ to the ejecta structures for the \snx fit in Figure \ref{f:lc_10x}. The Figure shows models with nickel masses of 0.01, 0.05, and 0.1 $\msun$; and for two levels of smearing, $s = $10 and 50. Less smearing (with nickel concentrated toward the center) is more likely to result in two distinct peaks.
 \label{f:lc_ni}}
\end{figure}

The contribution of significant emission from shock cooling does not necessarily preclude the presence of radioactive nickel in the ejecta. Models that include some radioactive $\nifs$ can produce more complex light curves with double-peaked morphologies. Figure \ref{f:lc_ni} shows our light curves using the parameters in Figure \ref{f:lc_10x} ($E_{\rm exp} = 3~B$, $\msh = 3~\msun$, $r_{\rm mid} = 2\times10^{12}~{\rm cm}$, and $\tau = 1~{\rm day}$) as well as 0.01, 0.05, or 0.1 $\msun$ of $\nifs$ concentrated in the center of the ejecta. The $\nifs$ is distributed throughout the ejecta using the parameterized radial profile  Equation \ref{eq:ni_dist} with smearing parameters $s = 10$ and $50$. These light curves qualitatively resemble those of double-peaked SNe discussed in \citet{drout16}, such as SNe 2005bf, 2008D, and 2013ge.

As expected, the additional nickel increases the peak luminosity and adds the characteristic radioactive tail. The $\nifs$ can also  produce a second peak in light curve,
but the radioactive peak can blend with the shock-cooling peak for models with smeared nickel distributions. Interestingly, the model with only $0.01~\msun$ of nickel but smearing factor $s = 50$ produces a bright, short-lived peak that drops precipitously to a very low magnitude, which might often be below the limits of detectors, depending on the object's distance. Therefore an object with a small amount of very smeared nickel in addition to the shock cooling contribution might increase the luminosity without producing a detectable tail.

\subsection{Effects of Rayleigh-Taylor Mixing \label{s:rt}}

While our hydrodynamical models have been carried out in 1D, it is well known that the SN interaction is subject to the Rayleigh-Taylor instability (RTI). 
 The sharp features and spikes in the density profiles of Figures~\ref{f:prof_vary_M} and \ref{f:prof_vary_tau} can be expected to smoothed out by RT instabilities, which will also mix the ejecta and CSM.
 These multi-dimensional affects could in principle affect the rate at which light diffuses out of the ejecta and could affect the shape of the light curve. 

\begin{figure}
\begin{center}
\includegraphics[width=3.5in]{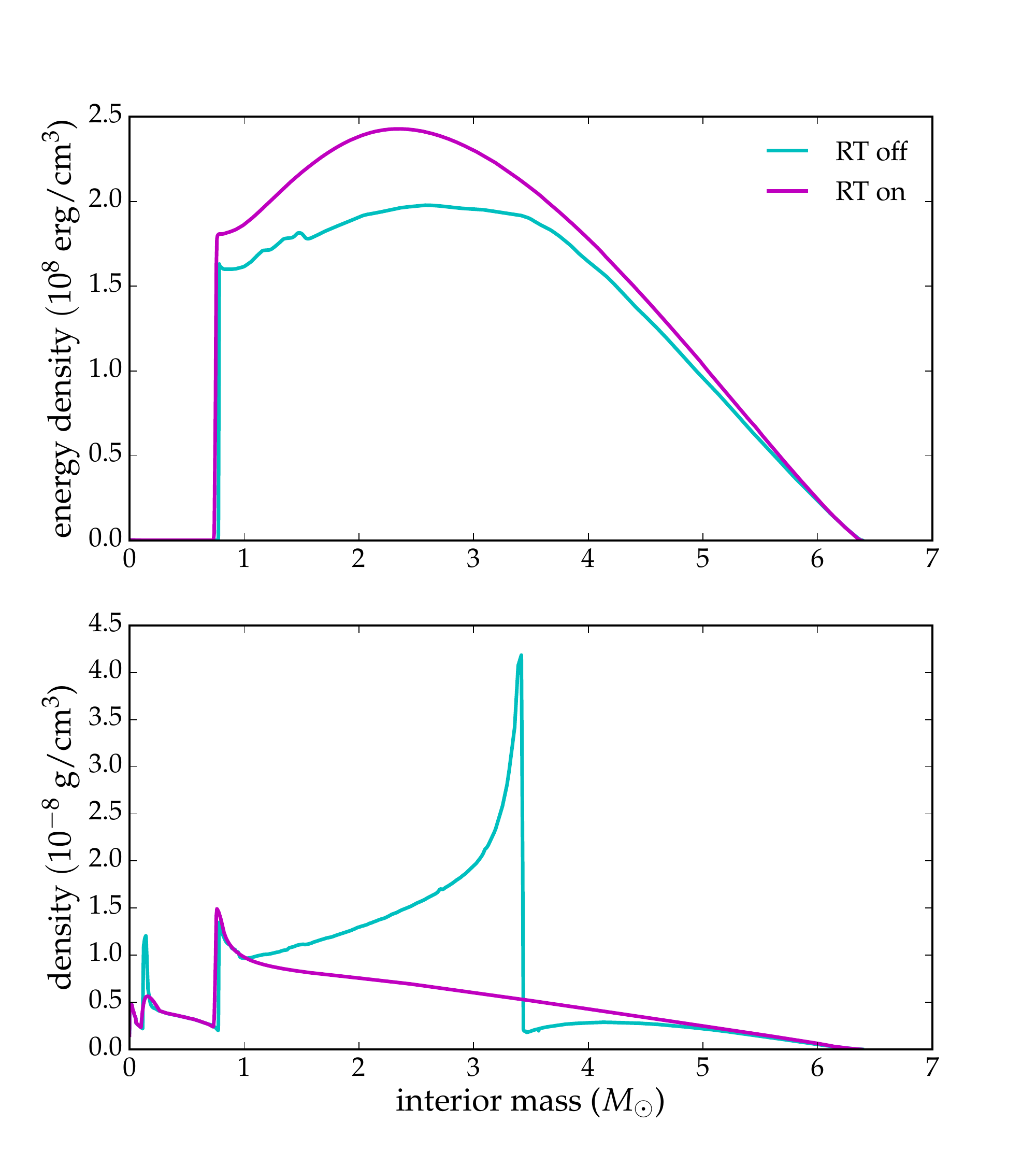}
\end{center}
\caption{Energy density and mass density profiles from the 1D hydrodynamics code from \citet{duffel16}, which includes a 3D-calibrated prescription for Rayleigh-Taylor mixing. Here the forward shock is stronger than shown in previous figures because we used a large radius ($2\times 10^{13}$ cm) in the hopes of capturing fast-rising, bright RFSNe. The density structure is dramatically affected by RT instabilities. Note that the run with Rayleigh-Taylor mixing on has a higher energy density; however the envelope is also not as extended as it is without mixing, since more of the outward kinetic energy is converted into turbulence. \label{f:rt_hydro}}
\end{figure}

To estimate the effects of the RTI on the models, we ran one of our star + CSM models using the hydrodynamics code from \citet{duffel16}, which includes a 1D RTI mixing prescription that has been calibrated to 3D models. In this case, we used a CSM mass of $3~\msun$ and a CSM radius of $2\times 10^{13}$ cm, chosen in order approach the higher luminosities of \snu and \snbj. The hydrodynamics results are shown in Figure \ref{f:rt_hydro}. RTI mixing almost entirely eliminates the large density spike that occurs in 1D models at the CSM/ejecta contact discontinuity. 
The energy density in the RTI calculation is also somewhat higher than a model without RTI, since kinetic energy in the form of turbulence eventually cascades into lower spatial scales until it is thermalized. Rather than all the kinetic energy go into expansion and acceleration of the ejecta, some instead becomes turbulent kinetic energy and eventually thermal energy.

\begin{figure}
\begin{center}
\includegraphics[width=3.5in]{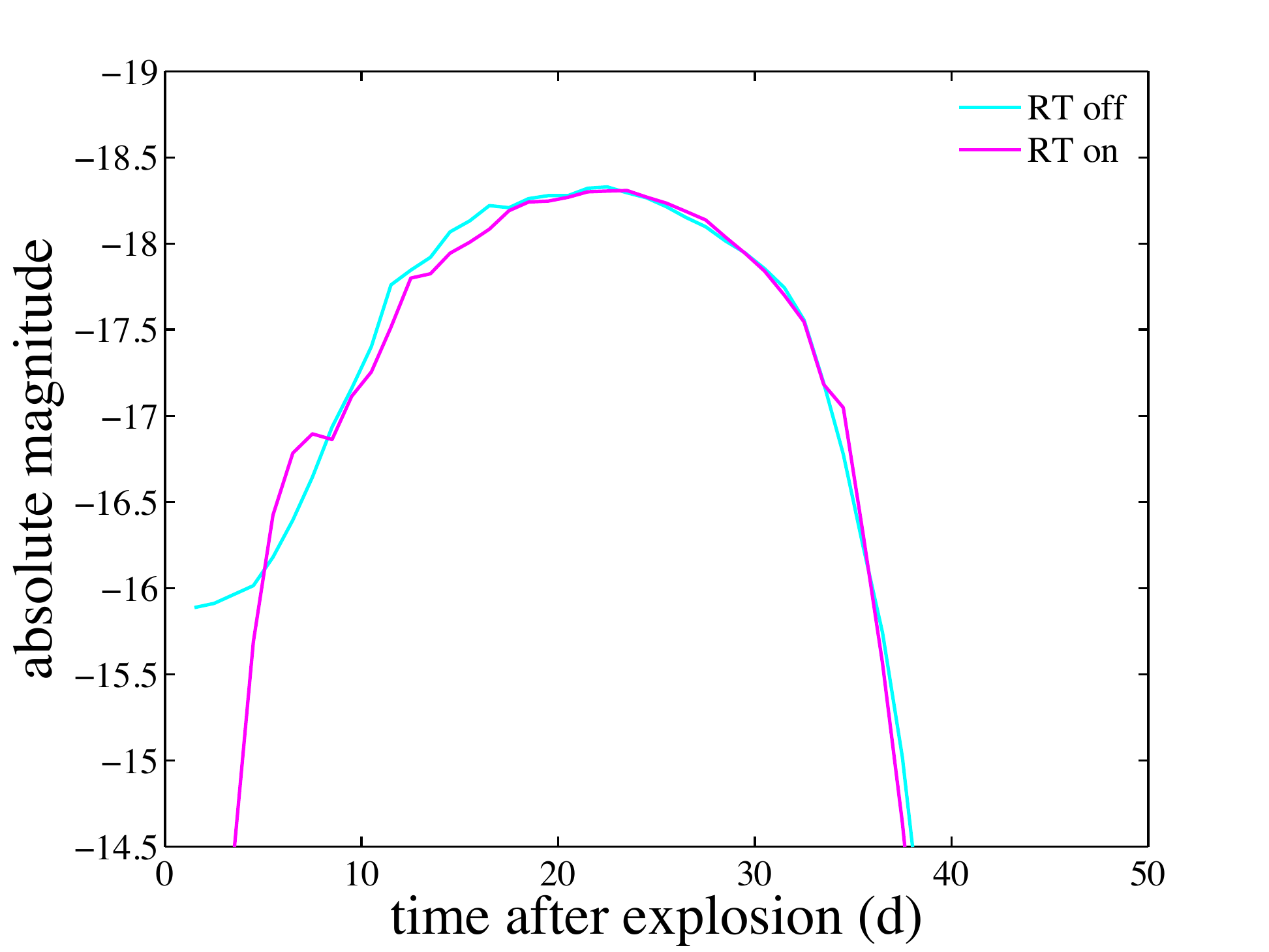}
\end{center}
\caption{Light curves using the hydro output from our code and the code from \citet{duffel16} with the Rayleigh-Taylor mixing prescription on and off. Evidently even though mixing can significantly affect the structure of the ejecta, it may not have a large effect on the bulk light curve properties.\label{f:rt_lc}}
\end{figure}

Figure \ref{f:rt_lc} shows the resulting light curves from the runs with RT prescription turned both on and off. It seems, in this case, that even though the final hydrodynamics profile is dramatically different, the mixing does not affect the overall peak luminosity or timescale, although it does affect the very early behavior of the light curve. This may be due to the fact that in the RT-off case, the shock passes through, heats, and accelerates the outer layers to large radii and large velocities, so the diffusion time for the small amount of radiation in these outer layers is short; in the RT-on case, much of the shock energy is dissipated into heat before it can reach these outer layers, and outer layers are not as accelerated and therefore do not reach the low densities needed for a very short diffusion time. In both runs, the peak luminosity is similar to that of \snbj, but the rise time is still too long to fit these fast-rising objects.

\section{Discussion and Future Directions} \label{s:discussion}

We have shown that models of the core-collapse SN with large pre-supernova radii and lacking $\nifs$ are a viable explanation for some H-free short-duration transients of a range of luminosities. We suggested that the large initial radius may be due to heavy mass loss just prior to the explosion, and we explored the dynamics and observable signatures of stars exploding into shells and winds.  The model light curves presented here resemble those of many of the observed RFSNe, but they struggle to capture the light curve shapes for some  objects with high luminosities and rapid rise times. It is likely that for brighter objects the stellar radius would be large enough that the shock has not propagated all the way through the shell by the time radiation losses become significant. Scenarios involving shock breakout in a wind may be more appropriate for these events, and this will be an area of exploration using radiation-hydrodynamical simulations in later work.
We expect that the use of radiation-hydrodynamics will change calculations for larger-radius progenitor systems. In such models, radiation will begin escaping at early times when the ejecta have not yet reached homologous expansion. These radiation losses can affect the dynamics; in particular, if radiation can escape directly from the region of the shock, the shock could lose significant energy and result is less acceleration of the outer layers. This could quantitatively change the peak and timescale of the light curve as well as the velocities of spectroscopic lines.

Two outstanding questions remain for the presented model  for RFSNe. One is the reason for the apparent low ejection of  $\nifs$.  Observations and parameterized 1D models of massive star explosions suggest that $\sim 0.05~\msun$ of $\nifs$ should be synthesized in typical core collapse events. In \S \ref{s:fallback}, we studied whether 
RFSNe may enhanced fallback, which could rob the ejecta of radioactivity. In stars surrounded by a dense CSM,  the interaction of the ejecta with the CSM will produce a reverse shock which can decelerate and push material back onto the central remnant. While this suggests an intriguing connection between nickel-free explosions and progenitors with extended envelopes or shells, achieving significant fallback through the reverse shock would require that the mass of the CSM more than exceed that of the ejecta.
Alternatively, independent of the presence of the CSM, fallback can occur if the explosion energy is somewhat less than the canonical 1~B.  We showed that for certain stellar structures, the explosion energy can be tuned to allow $\sim 0.1~\msun$ of
 of fallback while still unbinding the rest of the star and accelerating outer layers to high velocities. Light curves calculated for these examples are relatively dim and long-lived, so obtaining RFSNe with fallback may require lower-mass, higher-radius pre-SN configurations.
Our 1D studies, however, are merely a proof of concept for the viability of removing $\nifs$ by fallback. More detailed calculations would consider how the interior stellar structure may have been modified by the pre-supernova mass-loss, as well as the influence on fallback mass of both multi-dimensional dynamics and the particular explosion mechanism.

The second outstanding question is how H-stripped stars might be able to obtain extended envelopes or mass shell ejections that produce an adequately bright shock cooling light curve. While several theoretical studies have the laid the groundwork for understanding that late burning phases could unbind or extend much of the stellar envelope, more detailed stellar evolution calculations are needed to understand if these instabilities can occur in the final few days of a stripped envelope stars life.

\section*{Conclusions}
In this paper, we have explored the viability of hydrogen-stripped core-collapse supernova models using no radioactive nickel and extended helium envelopes to explain the enigmatic rapidly fading supernovae discovered in the last few years. Using 1D stellar evolution, hydrodynamics, and radiation transport codes in sequence, we have shown that such models reproduce the bulk properties of these events. We also compare our numerical results to analytical scalings predicted for the light curve properties. Further investigation using radiation-hydrodynamics codes would help understand the cases with more extended envelopes, as it is expected that sometimes the ejecta will still be dynamically interacting with the CSM even while radiation losses occur. Additional insight into possible mechanisms for both attaining such extended envelopes and failing to produce nickel in the ejecta are also necessary to validate this explanation.

\section*{Acknowledgements}
The authors would like to thank Sterl Phinney, Andrew McFadyen, Lars Bildsten, and Matteo Cantiello for useful discussion and collaboration. IK is supported by the DOE NNSA Stockpile Stewardship Graduate Fellowship Program. This research is also funded in part by the Gordon and Betty Moore Foundation through Grant GBMF5076. DK is supported in part by a Department of Energy Office of Nuclear Physics Early Career Award, and by the Director, Office of Energy Research, Office of High Energy and Nuclear Physics, Divisions of Nuclear Physics, of the U.S. Department of Energy under Contract No. DE-AC02-05CH11231.

\bibliographystyle{mn2e} 
\bibliography{truncated}

\end{document}